\documentclass[
prd,nofootinbib,longbibliography,a4paper,11pt,eqsecnum]{revtex4-1}
\usepackage{amsmath} 
\usepackage{graphicx} 
\usepackage{amsthm}
\usepackage{amssymb} 
\usepackage{physics}
\usepackage{dsfont}
\usepackage{yfonts}
\usepackage{hyperref}
\usepackage{graphicx}
\usepackage{epigraph}
\usepackage{slashed,stmaryrd}
\usepackage{soul}
\usepackage{subcaption}

\usepackage{array,xcolor,graphicx}
\usepackage{booktabs,multirow}
\usepackage[utf8]{inputenc}
\usepackage{mathtools}

\usepackage{etoolbox}
\patchcmd{\section}
  {\centering}
  {\raggedright}
  {}
  {}
\patchcmd{\subsection}
  {\centering}
  {\raggedright}
  {}
  {}
\usepackage[all]{xy}
\usepackage{tikz}
\usetikzlibrary{arrows.meta}

\hypersetup{colorlinks=true,linkcolor=black,citecolor=red,urlcolor=black}

\newcommand{\be}{\begin{equation}}
\newcommand{\ee}{\end{equation}}
\newcommand{\bea}{\begin{eqnarray}}
\newcommand{\eea}{\end{eqnarray}}

\def\la{\langle}
\def\ra{\rangle}
\def\Tr{{\mathrm{Tr}}}
\def\tr{{\mathrm{tr}}}

\def\d{\partial}

\def\CA{\mathcal{A}}

\def\CO{\mathcal{O}}

\def\Z{\mathbb{Z}}

\def\tT{\mathtt{T}}

\def\O{\text{O}}

\def\SU{\text{SU}}

\def\U{\text{U}(1)}

\def\Spin{\text{Spin}}
\def\Spinc{\text{Spin}_c}
\def\Pin{\text{Pin}}
\def\A{\mathcal{A}}

\def\UU{\text{U}}

\def\Tr{\text{~Tr~}}
\def\diag{\text{~diag}}
\def\ii{\text{i}}

\def\ee{\text{e}}
\def\dd{\text{d}}

\def\N{\mathbb{N}}

\def\hom{\text{~Hom}}
\def\ext{\text{Ext}}
\def\sq{\text{~Sq}}

\def\RPn (#1,#2){
  \fill (#1, #2) circle (3pt);
  \fill (#1, #2+1) circle (3pt);
}

\def\sqtwoL (#1,#2,#3){
  \draw[#3] (#1,#2) .. controls (#1-1,#2+1) .. (#1,#2+2);
}

\def\sqtwoR (#1,#2,#3){
  \draw[#3] (#1,#2) .. controls (#1+1,#2+1) .. (#1,#2+2);
}

\def \sqtwoCR (#1,#2,#3){
  \draw[#3] (#1,#2) .. controls (#1+1,#2+.5) and (#1+1.5,#2+2) .. (#1+2,#2+2);
}

\def \sqtwoCL (#1,#2,#3){
  \draw[#3] (#1,#2) .. controls (#1-1,#2+.5) and (#1-1.5,#2+2)  .. (#1-2,#2+2);
}

\def \sqone (#1,#2,#3){
  \draw[#3] (#1,#2) -- (#1,#2+1);
}

\def\Aone (#1,#2){
  \fill (#1, #2) circle (3pt);
  \fill (#1, #2+1) circle (3pt);
  \fill (#1, #2+2) circle (3pt);
  \fill (#1, #2+3) circle (3pt);
  \fill (#1+2, #2+3) circle (3pt);
  \fill (#1+2, #2+4) circle (3pt);
  \fill (#1+2, #2+5) circle (3pt);
  \fill (#1+2, #2+6) circle (3pt);
  \draw (#1, #2) -- (#1, #2+1);
  \draw (#1, #2+2) -- (#1, #2+3);
  \draw (#1+2, #2+3) -- (#1 + 2, #2+4);
  \draw (#1+2, #2+5) -- (#1+2, #2+6);
  \draw (#1, #2) .. controls (#1-1, #2+1) .. (#1, #2+2);
  \draw (#1+2, #2+4) .. controls (#1+3, #2+5) .. (#1+2, #2+6);
  \draw (#1, #2+1) .. controls (#1+1, #2+1.5) and  (#1+1.5 ,#2+3) .. (#1+2,#2+3);
  \draw (#1, #2+2) .. controls (#1+1, #2+2.5) and (#1+1.5, #2+4) .. (#1+2, #2+4);
  \draw (#1, #2+3) .. controls (#1+1, #2+3.5) and (#1+1.5, #2+5) .. (#1+2, #2+5);
}

\def\Aonecolor (#1,#2,#3){
  \fill[#3] (#1, #2) circle (3pt);
  \fill[#3] (#1, #2+1) circle (3pt);
  \fill[#3] (#1, #2+2) circle (3pt);
  \fill[#3] (#1, #2+3) circle (3pt);
  \fill[#3] (#1+2, #2+3) circle (3pt);
  \fill[#3] (#1+2, #2+4) circle (3pt);
  \fill[#3] (#1+2, #2+5) circle (3pt);
  \fill[#3] (#1+2, #2+6) circle (3pt);
  \draw[#3] (#1, #2) -- (#1, #2+1);
  \draw[#3] (#1, #2+2) -- (#1, #2+3);
  \draw[#3] (#1+2, #2+3) -- (#1 + 2, #2+4);
  \draw[#3] (#1+2, #2+5) -- (#1+2, #2+6);
  \draw[#3] (#1, #2) .. controls (#1-1, #2+1) .. (#1, #2+2);
  \draw[#3] (#1+2, #2+4) .. controls (#1+3, #2+5) .. (#1+2, #2+6);
  \draw[#3] (#1, #2+1) .. controls (#1+1, #2+1.5) and  (#1+1.5 ,#2+3) .. (#1+2,#2+3);
  \draw[#3] (#1, #2+2) .. controls (#1+1, #2+2.5) and (#1+1.5, #2+4) .. (#1+2, #2+4);
  \draw[#3] (#1, #2+3) .. controls (#1+1, #2+3.5) and (#1+1.5, #2+5) .. (#1+2, #2+5);
}

\def\rectangle (#1,#2,#3){   \draw[#3] (#1-0.15,#2-0.15) rectangle (#1+0.15,#2+0.15)}

\def\Eone (#1,#2){
  \fill (#1, #2) circle (3pt);
  \fill (#1, #2+1) circle (3pt);
  \fill (#1, #2+2) circle (3pt);
  \fill (#1, #2+3) circle (3pt);
  \draw (#1, #2) -- (#1, #2+1);
  \draw (#1, #2+2) -- (#1, #2+3);
  \draw (#1, #2) .. controls (#1-1, #2+1) .. (#1, #2+2);
  \draw (#1, #2+1) .. controls (#1+1, #2+2) .. (#1, #2+3);
}

\def\joker (#1,#2){
  \foreach \y in {#2, #2+1, #2+2, #2+3, #2+4}
  {\fill (#1,\y) circle (3pt);}
  \draw (#1,#2) -- (#1, #2+1);
  \draw (#1,#2+3) -- (#1, #2+4);
  \draw (#1,#2+0) .. controls (#1-1,#2+1) .. (#1, #2+2);
  \draw (#1,#2+2) .. controls (#1-1,#2+3) .. (#1, #2+4);
  \draw (#1,#2+1) .. controls (#1+1,#2+2) .. (#1, #2+3);
}

\def\jokercolor (#1,#2, #3){
  \foreach \y in {#2, #2+1, #2+2, #2+3, #2+4}
  {\fill[#3] (#1,\y) circle (3pt);}
  \draw[#3] (#1,#2) -- (#1, #2+1);
  \draw[#3] (#1,#2+3) -- (#1, #2+4);
  \draw[#3] (#1,#2+0) .. controls (#1-1,#2+1) .. (#1, #2+2);
  \draw[#3] (#1,#2+2) .. controls (#1-1,#2+3) .. (#1, #2+4);
  \draw[#3] (#1,#2+1) .. controls (#1+1,#2+2) .. (#1, #2+3);
}

\def\msopart (#1,#2,#3){
  \fill[#3] (#1,#2) circle (3pt); 
  \fill[#3] (#1, #2+1) circle (3pt);
  \fill[#3] (#1, #2+2) circle (3pt);
  \fill[#3] (#1+2, #2+2) circle (3pt);
  \fill[#3] (#1+2, #2+3) circle (3pt);
  \fill[#3] (#1+2, #2+4) circle (3pt);
  \fill[#3] (#1+2, #2+5) circle (3pt);
  \sqtwoCR(#1,#2, #3);
  \sqtwoCR (#1, #2+1, #3);
  \sqtwoCR (#1, #2+2, #3);
  \sqone (#1+2, #2+2, #3);
  \sqone (#1+2, #2+4, #3);
  \sqtwoR(#1+2, #2+3, #3);
  \sqone (#1, #2+1, #3); }

\def\amme (#1,#2,#3){
  \fill[#3] (#1,#2) circle (3pt) ;
  \sqtwoR(#1,#2,#3);
  \fill[#3] (#1,#2+2) circle (3pt) ;
  \sqone(#1,#2+2,#3);
  \fill[#3] (#1,#2+3) circle (3pt) ;
  \sqtwoR(#1,#2+3,#3);
  \fill[#3] (#1,#2+5) circle (3pt) ;}

\def\questionupsidedon (#1,#2,#3){
  \fill[#3] (#1,#2) circle (3pt) ;
  \sqtwoR(#1,#2,#3);
  \fill[#3] (#1,#2+2) circle (3pt) ;
  \sqone(#1,#2+2,#3);
  \fill[#3] (#1,#2+3) circle (3pt) ;}

\def\emm (#1,#2,#3,#4){
  \fill[#3] (#1,#2) circle (3pt) node[anchor=east] {$#4$};
  \fill[#3] (#1,#2+1) circle (3pt);
  \fill[#3] (#1,#2+2) circle (3pt);
  \fill[#3] (#1,#2+3) circle (3pt);
  \fill[#3] (#1,#2+4) circle (3pt);
  \fill[#3] (#1,#2+5) circle (3pt);
  \fill[#3] (#1,#2+6) circle (3pt);
  \fill[#3] (#1,#2+7) circle (3pt);
  \fill[#3] (#1,#2+8) circle (3pt);

  \sqone (#1,#2,#3);
  \sqtwoR (#1,#2+1,#3);
  \sqone (#1,#2+2,#3);
  \sqtwoL (#1,#2+2,#3);
  \sqone (#1,#2+4,#3);
  \sqtwoL (#1,#2+5,#3);
  \sqone (#1,#2+6,#3);
  \sqtwoR (#1,#2+6,#3);

  \draw[dashed,#3] (#1,#2+8) -- (#1,#2+9);
}

\def\Minf (#1,#2,#3){
  \fill[#3] (#1,#2) circle (3pt);
  \fill[#3] (#1,#2+2) circle (3pt);
  \fill[#3] (#1,#2+3) circle (3pt);
  \fill[#3] (#1,#2+4) circle (3pt);
  \fill[#3] (#1,#2+5) circle (3pt);
  \fill[#3] (#1,#2+6) circle (3pt);

  \sqtwoL (#1,#2,#3);
  \sqone (#1,#2+2,#3);
  \sqtwoL (#1,#2+3,#3);
  \sqone (#1,#2+4,#3);
  \sqtwoR (#1,#2+4,#3);

  \draw[dashed,#3] (#1,#2+6) -- (#1,#2+7);
}

\begin{document}

\rightline{CERN-TH-2023-213}

\bigskip
    
\title{A Non-Perturbative Mixed Anomaly \\ and Fractional Hydrodynamic Transport 
}

\author{Joe Davighi}
\affiliation{Theoretical Physics Department, CERN, 1211 Geneva 23, Switzerland}
\email{joseph.davighi@cern.ch}

\author{Nakarin Lohitsiri}
\affiliation{Department of Mathematical Sciences, Durham University,
Upper Mountjoy, Stockton Road, Durham, DH1 3LE, United Kingdom}
\email{nakarin.lohitsiri@durham.ac.uk}

\author{Napat Poovuttikul}
\affiliation{High Energy Physics Research Unit, Department of Physics,
Faculty of Science, Chulalongkorn University, Bangkok 10330, Thailand}
\email{napat.po@chula.ac.th}

\begin{abstract}
\vspace{1cm}

We present a new non-perturbative 't Hooft anomaly afflicting a quantum field theory with symmetry group $G=\U\times \mathbb{Z}_2$ in four dimensions. We use the Adams spectral sequence to compute that the bordism group $\Omega^{\Spin}_5(BG)$, which classifies anomalies that remain when perturbative anomalies cancel, is $\mathbb{Z}_4$. By constructing a mapping torus and evaluating the Atiyah--Patodi--Singer $\eta$-invariant, we show that the mod 4 anomaly is generated by a pair of Weyl fermions that are vector-like under $\U$, but with only one component charged under $\mathbb{Z}_2$. We construct a simple microscopic field theory that realises the anomaly, before investigating its impact in the hydrodynamic limit. We find that the anomaly dictates transport phenomena in the $\U$ current and energy-momentum tensor akin to the chiral vortical and magnetic effects (even though the perturbative anomalies here vanish), but with the conductivities being fractionally quantised in units of a quarter, reflecting the mod 4 nature of the bordism group. 
Along the way, we compute the (relevant) bordism groups $\Omega^{\Spin}_d(B\mathbb{Z}_2\times B\U)$ and $\Omega_d^{\Pin^-}(B\U)$ in all degrees $d=0$ through 5.

\end{abstract}
\maketitle

\tableofcontents

\section{Introduction}

An anomaly is, generally speaking, some kind of obstruction to quantisation. Usually (but not always~\cite{Freed:2023snr}) we are concerned with anomalous {\em symmetries}. At least in the case of chiral symmetries acting on massless fermions, the anomaly is then related to a topological index that requires its coefficient be integer-quantised. As a result, this type of anomaly cannot change under the renormalization group (RG) flow of a theory, which is a continuous deformation. 
The anomaly therefore communicates robust information from microscopic scales to macroscopic scales. 
This idea of {\em anomaly matching} over different scales has been immensely powerful in (a) constraining the infrared dynamics of quantum field theories (QFTs), even when confronted by strong coupling \cite{tHooft:1979rat}; (b) providing highly non-trivial consistency checks on hypothesized dualities {\em e.g.}~\cite{Seiberg:1994pq,Intriligator:1995id,Pouliot:1995zc,Pouliot:1995sk}; and (c) explaining particular physical phenomena, most famously the observed neutral pion decay to photons \cite{Adler:1969gk, Bell:1969ts}. 

While in high-energy physics examples such as these we are accustomed to applying the anomaly matching to QFTs at zero temperature, the principle applies just as well for theories at finite temperature and/or finite chemical potential. At least for microscopic theories that are both gapped and interacting, the {\em hydrodynamic limit} obtained by coarse-graining over all particle degrees of freedom arguably provides an ultimate `effective theory', in which all the dynamics are captured by an effective action for the background fields associated to conserved currents for the global symmetries of the microscopic theory. Said more prosaically, all that remains of the microscopic physics in the hydrodynamic limit are its symmetries. The symmetry currents of interest typically include a stress-energy tensor $T^{\mu\nu}$, and currents $j^\mu$ associated {\em e.g.} with $\U$ particle number symmetry, for which background fields are a spacetime metric $g_{\mu\nu}$ and gauge field $A_\mu$ respectively. 
Anomaly matching is now known to play a powerful role in this hydrodynamic context \cite{Vilenkin:1980fu,Alekseev:1998ds,Son2009,Neiman:2010zi,Landsteiner:2011cp,Banerjee:2012iz,Jensen:2012kj}, at least for perturbative 't Hooft anomalies, by constraining the coefficients of Chern--Simons-like terms appearing in the low-energy effective action. These topological terms in the effective action manifest themselves physically via certain {\em transport coefficients}. For example, an anomalous microscopic $\U$ symmetry might imply there is momentum and/or $\U$ flux flowing in the direction of magnetic fields and/or vorticity, as we will review in \S \ref{sec:hydro-review}.


In recent years, our understanding of anomalies has become more mathematically rigorous. This is rooted in the idea of {\em anomaly inflow}~\cite{Callan:1984sa} -- which has a physical incarnation in condensed matter phenomena such as the quantum Hall effect -- whereby anomalies in $d$-dimensional QFTs are captured by extending the theory to a bulk spacetime in $d+1$ dimensions, whose boundary is the original spacetime. The anomaly is itself identified with a QFT in $d+1$ dimensions, with special properties. In particular, because the anomalous transformation always induces a {\em re-phasing} of the partition function, the anomaly theory is an {\em invertible} QFT. A pioneering work of Freed and Hopkins~\cite{Freed:2016rqq} showed how such unitary, invertible QFTs can be classified algebraically, using a particular cohomology theory called {\em cobordism}.\footnote{Technically speaking, cobordism here refers to the shifted Anderson dual of bordism, where the latter is a homology theory whose elements are equivalence classes of manifolds (equipped with structures, like gauge bundles for the symmetry group) that can be `connected' by an interpolating manifold in one higher dimension.}
Our particular interest here is in chiral symmetries of massless fermions in $d$ spacetime dimensions, so let us now recap in a more explicit manner how the cobordism classification of anomalies emerges in this context. 

The key object is the fermionic partition function, formally the path integral over chiral fermion fields $Z[A]=\int \mathcal{D}\psi \mathcal{D}\bar{\psi} e^{i S[\psi,A]}$, which is a functional of the background gauge fields that we collectively denote $A$. 
At least for fermions weakly coupled to a gauge field via the usual kinetic term, this partition function is the determinant of a suitably defined Dirac operator. Witten and Yonekura recently proved~\cite{Witten:2019bou}, building on {\em e.g.} Refs.~\cite{Witten:1985xe,Dai:1994kq,Witten:2015aba}, that the phase of this fermion partition function is the exponentiated $\eta$-invariant appearing in the index theorem of Atiyah, Patodi, and Singer~\cite{Atiyah:1975jf,Atiyah:1976jg,Atiyah:1976qjr}, evaluated on a bulk $(d+1)$-manifold whose boundary is the original spacetime \`a la anomaly inflow. 
Equipped with this formula for the phase of $Z[A]$, one can then compute how that phase shifts under any transformation of the background fields, which can involve both gauge transformations $A \to A^g$ but also diffeomorphisms of the background geometry, by evaluating $\exp(-2\pi i \eta)$ on a $(d+1)$-dimensional {\em mapping torus} that interpolates between the two field configurations.\footnote{
Note that our convention for the $\eta$-invariant, particularly in relation to bordism group computations, follows from Ref. \cite{Witten:2019bou}, which differs from the convention used in \cite{Golkar:2015oxw,Chowdhury:2016cmh} by a factor of 2.
} Such mapping tori probe all possible anomalies, and so the original QFT is completely anomaly-free if $\exp(-2\pi i \eta) = 1$ on all possible mapping tori. 

In fact, requiring that $\exp(-2\pi i \eta) = 1$ on all closed $(d+1)$-manifolds equipped with all possible background field configurations (not just tori) is motivated by demanding {\em locality} of the fermion partition function; in this way, we can understand anomaly cancellation to follow necessarily from locality. This `strong' notion of locality is thought to be an important ingredient in embedding our QFT in a consistent description of quantum gravity, as suggested in {\em e.g.}~\cite{Garcia-Etxebarria:2018ajm}.

To summarise, computing $\eta$-invariants detects all possible chiral fermion anomalies. In perturbation theory, one can always choose a mapping torus that is itself the boundary of a $(d+2)$-manifold $Y$, and thus use the APS index theorem to learn that $Z[A^g] = Z[A] \exp(-2\pi i\int_Y \Phi_{d+2})$, where $\Phi_{d+2}$ is the {\em anomaly polynomial} appearing in the index theorem (for which we will see a precise formula later). This object detects all possible perturbative ({\em a.k.a.} `local') anomalies. But the real power of the anomaly inflow formula lies in classifying {\em non-perturbative}, or `global' anomalies. A canonical example of such a global anomaly is that afflicting a 4d $\SU(2)$ gauge theory with an odd number of Weyl doublets~\cite{Witten:1982fp}. When local anomalies vanish ($\Phi_{d+2}=0$), the APS index theorem tells us that $\exp(-2\pi i \eta)$, which detects all possible remaining anomalies when evaluated on mapping tori, is a {\em bordism invariant}, which means that it is the trivial phase when evaluated on a manifold that is a boundary.
The global anomaly is therefore a homomorphism from the bordism group $\Omega_{d+1}$ to $\U$; we `just' need to compute it on each generator of the bordism group to determine all possible anomalous phases that can arise for such a symmetry. This information is very powerful. For instance, if the appropriate bordism group vanishes, we immediately learn that there can be no global anomaly associated with any symmetry transformation, for any spacetime geometry and background field configuration \cite{Freed:2006mx,Garcia-Etxebarria:2017crf,Garcia-Etxebarria:2018ajm,Wan:2018bns,Seiberg:2018ntt,Hsieh:2019iba,Davighi:2019rcd,Wan:2019gqr,Davighi:2020uab,Lee:2020ewl,Davighi:2020kok,Debray:2021vob,Davighi:2022fer,Davighi:2022bqf,Wang:2022eag,Lee:2022spd,Davighi:2022icj,Debray:2023yrs,Basile:2023knk}.

It is not {\em a priori} obvious whether global anomalies of this kind should have any physical consequences in the hydrodynamic limit of a theory.\footnote{We remark that the hydrodynamic consequences of an anomalous `large symmetry transformation' was considered in Ref.~\cite{Golkar:2015oxw}, which is close in spirit to the present work. The symmetry type considered there is $\Spin(4) \times \U$, in the 4d case, and the symmetry transformation involves turning on a $\U$ monopole background and doing a large diffeomorphism. The anomaly is studied by computing the $\eta$-invariant. Nonetheless, that anomaly comes from a term in the anomaly polynomial, namely $ \frac{1}{24} p_1(R) c_1(F) \subset \Phi_6$, and so should be classified as a perturbative anomaly. Indeed, that anomaly can already be seen by doing an infinitesimal gauge transformation; in particular, by taking a gravitational instanton background, for example on the K3 surface (which has non-vanishing signature), and doing any infinitesimal $\U$ gauge transformation. Once this perturbative anomaly has been cancelled, which requires the sum of $\U$ charges vanishes, there is no remaining anomaly associated with any large symmetry transformation. This can be verified from the vanishing of the bordism group $\Omega_5^\Spin(B\U)$~\cite{Garcia-Etxebarria:2018ajm}. \label{foot:Golkar-Sethi}
} 
The equilibrium partition function for the fluid is, as described above, a functional of the background fields for the (continuum) global symmetries that characterise the theory, such as $g_{\mu\nu}$ and a $\U$ background $A_\mu$. In contrast to the case of local anomalies, which recall can be matched by Chern--Simons forms built from the Riemann curvature $R$ and the field strength $F$, the low-energy effective action that is needed to match a global anomaly cannot be captured using only this data. It might therefore appear that a global anomaly can play no role. 

In this paper, we use the cobordism classification to identify a novel {\em mixed} global anomaly for a 4d microscopic theory with $\Spin \times \U \times \Z_2$ global symmetry, before exploring its consequences in the hydrodynamic limit.
While neither $\U$ nor $\Z_2$ can carry a global anomaly on its own in 4d, we find that the presence of both symmetries allows for a global anomaly.\footnote{A qualitatively similar anomaly, probed by the bordism group $\Omega_5^{\Spinc}(B\Z_2) \cong \Z_8 \times \Z_2$, was very recently studied by Chen, Hsieh, and Matsudo in Ref.~\cite{Chen:2023hmm} in connection with the fermion-monopole scattering problem~\cite{vanBeest:2023dbu} and related reflection anomalies in 2+1 dimensions.}
The 't Hooft anomaly itself is valued in $\Z_4$, and is generated by a pair of Weyl fermions that are vector-like under the $\U$, but for which only one Weyl component is charged under the $\Z_2$. This symmetry type, and the anomalous fermion content, arises in very simple microscopic theories; for example, in which Weyl fermions are coupled to massive real scalars. We explicitly construct a mapping torus that probes this anomaly, which involves doing a large diffeomorphism in the presence of both a non-trivial $\Z_2$ holonomy and non-zero $\U$ monopole flux -- the necessity of both these things is not surprising, and confirms the assertion that it is a mixed anomaly. The fact that our mapping torus does provide a valid generator for the bordism group $\Omega_5^\Spin(B(\U \times \Z_2))=\Z_4$ is verified by an explicit calculation of the $\eta$-invariant thereon, which we do using the idea of `anomaly interplay' as introduced in Refs.~\cite{Davighi:2020bvi,Davighi:2020uab}.

We then turn to the hydrodynamic limit of such a microscopic theory, at finite temperature and chemical potential, to explore the consequences of the mixed anomaly. The $\U$ symmetry that participates in the 't Hooft anomaly has a conserved current $j^\mu$, obtained by varying the equilibrium partition function $Z$ with respect to the $\U$ background gauge field $A_\mu$ as usual. We find that, if there is $\Z_2$ holonomy around the time-circle on which we compactify the theory at finite $T$, the mixed anomaly leaves its mark in the effective field theory. Even though there is no perturbative anomaly in the 4d microscopic theory, matching the mixed global anomaly requires there to be non-trivial 3d Chern--Simons terms on spatial hypersurfaces, whose coefficients are {\em fractionally quantised} in units of a quarter. Such a fractional Chern--Simons level is itself a hallmark of a global anomaly; being fractional, it cannot be removed by a local counterterm -- unless, of course, the global anomaly vanishes.

The constraints that we derive at the level of the effective action translate directly to constraints on certain transport coefficients for both $T^{\mu\nu}$ and $j^\mu$, meaning the anomaly, while subtle, has dramatic physical implications. We show that the Noether currents $j^\mu,T^{\mu\nu}$ are compelled to have a non-zero component parallel to the magnetic field and/or fluid vorticity $\omega^\mu := \epsilon^{\mu\nu\rho\sigma}u_\nu \partial_\rho u_\sigma$ when the mixed anomaly is non-vanishing. This phenomena is a non-perturbative anomaly version of the well-known chiral magnetic and chiral vortical effects of \cite{Fukushima:2008xe} and, to the best of our knowledge, is the first such example. We confirm the results of our anomaly-matching arguments by explicitly calculating the transport coefficients using the associated Kubo formulae, which relate them to the IR limits of 2-point functions in the finite temperature QFT, for a microscopic theory of free fermions (that exhibit the 1 mod 4 valued 't Hooft anomaly in question). The key calculational step here, by which the $\Z_2$ symmetry enters the picture, is that the presence of $\Z_2$ holonomy dictates a certain choice of non-standard boundary conditions for our free fermion wavefunctions.

We emphasise that, if the discrete $\Z_2$ global symmetry were neglected, then we would predict this transport coefficient (and others) to be zero, again by anomaly matching, because there is no perturbative anomaly. This phenomenon uncovers a surprising way in which a discrete symmetry in a microscopic theory can leave its vestige in the hydrodynamic limit, even though there is no associated continuous current for that symmetry appearing in the fluid equations of motion. The long-distance transport behaviour only becomes sensitive to the presence of a microscopic discrete symmetry when there is the global mixed anomaly. This exemplifies the power of anomaly-matching in effective field theory.

The structure of the rest of the paper is as follows. In \S \ref{sec:hydro-review}, we briefly review the equilibrium constraints on hydrodynamics and the effective partition function construction. Our purpose here is to introduce key ingredients required to describe the effect of the non-perturbative anomaly, as well as to clarify differences between this and the well-known effects from perturbative anomalies. In \S \ref{sec:formal-stuff}, we derive the $\U\times \mathbb{Z}_2$ anomaly using bordism theory. We explicitly construct a 5-dimensional mapping torus that probes the anomaly and compute the $\eta$-invariant thereon, and we present a  microscopic realisation of this anomaly. \S \ref{sec:transport} presents the result of anomaly-induced transport coefficients associated to this mixed anomaly, firstly via a general macroscopic anomaly-matching argument, which we then verify in a particular free fermion limit by direct computation. We summarise our results and discuss possible future directions in \S \ref{sec:conclusion}. We present bordism group computations for $\Omega^{\Spin}_d(B(\U\times \Z_2))$, as well as for related symmetry structures $\Omega_d^{\Pin^-}(B\U)$ and $\Omega_d^{\mathrm{Spin}_c}(B\mathbb{Z}_2)$,\footnote{The particular case of $\Omega_d^{\mathrm{Spin}_c}(B\mathbb{Z}_2)$ bordism groups have been already calculated by other more analytical methods, going back to classic computations of Bahri and Gilkey in Ref.~\cite{bahri1987eta}. } 
using the Adams spectral sequence, together with a short introduction to the method, in Appendix \ref{app:bordism}.

\section{Hydrodynamics: partition functions and anomaly matching} \label{sec:hydro-review}

Relativistic hydrodynamics is often viewed as a system of differential equations, consisting of the Ward identities for continuous global symmetries. In the case considered in this work, these are translational and $\U$ symmetries, with the associated Noether currents $T^{\mu\nu}$ and $j^\mu$, respectively. The system is said to be in the normal phase of hydrodynamics when these operators can be expressed in terms of the local densities of the conserved charges and/or their conjugates, such as the temperature $T$, the $\U$ chemical potential $\mu$, and the velocity $u^\mu$, as well as the background fields. The Noether currents are then constructed order by order in the gradient expansions. These expansions are referred to as the constitutive relations:   
\begin{equation} 
T^{\mu\nu} = (\varepsilon+ p) u^\mu u^\nu + p g^{\mu\nu} + T^{\mu\nu}_\text{1st}+ \CO(\d^2) \,, \qquad j^\mu = n u^\mu + j^\mu_\text{1st}  + \CO(\d^2) \, , \qquad 
\end{equation}
where $T^{\mu \nu}_\text{1st}$ and $j^\mu_\text{1st}$ are sums of all possible symmetric rank-2 tensors and vectors, respectively, at first order in the derivative expansions.\footnote{
For example, $T^{\mu\nu}_\text{1st} \supset -\eta \sigma^{\mu\nu}$, where $\sigma^{\mu\nu}$ is the shear tensor constructed from the symmetric-traceless part of $\d_\mu u_\nu$. This is a standard approach elucidated in {\em e.g.} \cite{LLfluid}. See also \cite{Kovtun:2012rj} for a modern review on the subject.
}
The coefficients accompanying each structure, known as {\em transport coefficients} for the hydrodynamic system, are not arbitrary -- they are determined by the microscopic theory that underlies the fluid description. However, even without access to the microscopic picture, there are macroscopic consistency conditions that these coefficients must satisfy. Chief among them is that when the system is in equilibrium, the Noether currents must reduce to those obtained from an equilibrium partition function via
\begin{equation}
    T^{\mu\nu} =-\frac{i}{\sqrt{-g}}\frac{\delta }{\delta g_{\mu\nu}} \log Z\,, \qquad     j^{\mu} =-\frac{i}{\sqrt{-g}}\frac{\delta }{\delta a_{\mu}} \log Z\, ,
\end{equation}
where $Z$ is the analytic continuation of the thermal partition function with all the background fields turns on. This condition has been explored in various kinds of fluid and has been shown to put strong constraints on various transport coefficients \cite{Banerjee:2012iz,Jensen:2012jh}.\footnote{
For example, applying this constraint at the zeroth derivative level, one finds that $p$ is the density of $\log Z$ and relations 
\begin{equation}
    d p = s dT + n d\mu \, , \quad\text{and}\quad \varepsilon+ p = sT + \mu n\, ,
\end{equation}
which are the first law of thermodynamics and the extensivity condition respectively. 
} 

Let us briefly review the systematic construction of the equilibrium partition function based on the global symmetry of the system. 
%
%
%
Since we are interested in describing hydrodynamic systems in equilibrium, we consider only configurations that are invariant under time translations.
This means the background metric $g$ admits a time-like Killing vector $K_t$ ({\em i.e.} the Lie derivative $\mathcal{L}_{K_t} g=0$) which, without loss of generality, we can take to be $K_t=\partial/\partial t$ by choice of our time-like coordinate $t$. The most general geometry on a 4-manifold $X_4$ that admits such a Killing vector is an $S^1$ fibration over a (spatial) manifold $X_3$, which can be parametrized via the Kaluza--Klein ansatz, as follows (see {\em e.g.}~\cite{Banerjee:2012iz}):
\begin{equation}
    g = e^{2\sigma(x)} (dt + \alpha_i(x) dx^i)^2 + \gamma_{ij}(x) dx^i dx^j\, ,
\end{equation}
where $x^i$ are coordinates parametrizing the spatial hypersurfaces normal to $K_t$. Beyond the time-translation symmetry we imposed, any such metric admits more general symmetries of the form:
\begin{equation} \label{eq:time-shift}
    t \mapsto t + \phi(x), \qquad \alpha_i(x) \mapsto \alpha_i(x) - \partial_i \phi(x)\, ,
\end{equation}
which is formally equivalent to doing a 3d $\U$ gauge transformation on $\alpha_i$, regarded as the components of a dimensionally-reduced $\U$  connection. It is conventional to refer to this symmetry as a `Kaluza--Klein (KK) gauge transformation'.

We now couple the theory to a conserved $\U$ current, for which we introduce a background gauge field $a = a_t dt + a_i dx^i$.
In doing so we wish to preserve time-translation symmetry, meaning we require $\mathcal{L}_{K_t} a = 0$, which simply implies that the components $a_t$ and $a_i$ are functions only of $x^i$. The $\U$ gauge transformations of interest act as
\begin{equation}
a_t(x) \mapsto a_t(x), \qquad
a_i(x) \mapsto a_i(x) + \partial_i \lambda(x)\, ,
\end{equation}
where again we do not consider gauge transformations that would take us to configurations with an explicit time-dependence; this restriction can be thought of as a partial gauge-fixing.
%
It is then convenient to split up the $\U$ connection $a$ into terms that are each invariant under the transformation (\ref{eq:time-shift}). We decompose
\begin{equation}
    a = a_t(x) (dt+ \alpha_i(x) dx^i) + \mathcal{A}_i(x) dx^i\, .
\end{equation}
Since the combination $dt+\alpha_i dx^i$ is invariant under (\ref{eq:time-shift}), it follows that the object $\mathcal{A}_i$ is also invariant under the KK gauge transformation.

The fluid variables can be expressed in terms of the background fields as follows. Starting from the zeroth derivative quantities, we have
\begin{equation}
u^\mu = ( e^{-\sigma},0,0,0) \, , \qquad T = \beta^{-1}e^{-\sigma}\, , \qquad \mu = e^{-\sigma} a_t
\end{equation}
up to the derivative corrections. Almost all first derivative structures vanishes in this configuration except the magnetic field $B^\mu$ and the vorticity $\omega^\mu$,
\begin{equation}
B^\mu := \frac{1}{2}\epsilon^{\mu \nu \rho \sigma} u_\nu (da)_{\rho \sigma}\, ,
\qquad
\omega^\mu := \epsilon^{\mu \nu \rho \sigma} u_\nu \d_\rho u_\sigma\, , 
\end{equation}
which can be related to the field strengths of the two abelian gauge fields via
\begin{equation}
	(d \alpha)_{\mu \nu} = \frac{1}{2}\epsilon_{\mu \nu \rho \sigma}u^\rho \omega^\sigma \, , \qquad (d\CA)_{\mu \nu} = \frac{1}{2} \epsilon_{\mu \nu \rho \sigma}  u^\rho ( B^\sigma + \mu \omega^\sigma )\, .
\end{equation}
With all these ingredients, one can write down the most general equilibrium partition function $Z[g,a] = Z[e^{\sigma}, \alpha_i, \gamma_{ij}, a_t,\CA_i]$ that is invariant under both diffeomorphisms and gauge transformations, to ensure the conservation of the Noether currents. Since we are interested in the long-wavelength limit, where the thermal size $\beta \sim 1/T$ is small compared to the gradient of the hydrodynamic variables, we can expand $W= -\log Z$ order by order in the gradient expansion. Expressed in the terms of the dimensionally reduced 3d effective action on $X_3$, the leading terms are 
\begin{equation}\label{eq:genericW}
    W= -\log Z = \beta \int_{X_3} \star_3(e^{\sigma} p[T,\mu]) + i\int_{X_3} c_0 \CA\wedge d\CA + i\beta^{-2} \int_{X_3} c_1 \alpha\wedge d\alpha + i\beta^{-1} \int_{X_3} c_2\alpha \wedge d\CA + \CO(\d^2)
\end{equation}
where $\star_3(\cdot) $ denotes the hodge dual of a 0-form on the three-dimensional spatial slice $X_3$, which is assumed to have no boundary. 

The nature of the coefficients $c_0$, $c_1$ and $c_2$ determines the physical consequences of these first derivative terms, as follows:    \begin{itemize}
	\item If the $c_i$ are constants and properly quantised, namely 
	\begin{equation}\label{eq:c0c1c2-contactTerm}
		c_0 = \frac{n}{4\pi} \, , \qquad c_1,c_2 = \frac{n}{2\pi}\,,\qquad n \in \mathbb{Z},
	\end{equation}
	then these are {\em bona fide} Chern--Simons term for the background gauge fields. They are gauge invariant contributions to the exponentiated effective action, and their presence preserves both the $\U$ symmetry and diffeomorphism invariance. However, these are contact terms and as such they can be added or removed from this system. In other words one can, in principle, redefine the microscopic theory with the contact term included, and thereby freely shift the coefficients of this kind in the effective description.
 Furthermore, by requiring that the action $W$ is invariant under CPT symmetry, one notices that the pure Chern--Simons terms are CPT odd. Thus the coefficients $c_0,c_1$ have to vanish, and only the mixed Chern--Simons coefficient $c_2$ can remain non-zero \cite{Banerjee:2012iz}. 
 
 
\item If the $c_i$ are not constants, then the system loses the global symmetry. Interestingly, when the non-invariance take a particular form, it can be used to capture the non-conservation of currents in theories with perturbative anomalies, as we shall soon recap. See {\em e.g.} Refs.~\cite{Banerjee:2012iz,Banerjee:2012cr,Jensen:2012kj,Jensen:2013rga}.
\item Finally, if the $c_i$ are constant but not quantised as in \eqref{eq:c0c1c2-contactTerm}, then the effective action will typically transform under `large' gauge transformations. The integer-quantised part of it can be removed by a contact term, but the fractional part is physical and reflects the global anomaly of the system. The case considered in the present paper is of this type. More discussion about this type of contact term in 2+1 dimensions can be found in Ref.~\cite{Closset:2012vp}, and an earlier application to hydrodynamics in Ref.~\cite{Poovuttikul:2021mxx}.
\end{itemize}
These Chern--Simons terms have a profound effect on transport phenomena in the hydrodynamic limit, as captured by the {\em transport coefficients} $\lambda_{1,2}$, $\zeta_{1,2}$ defined via the constitutive relations: 
\begin{equation}\label{eq:1stOrder-constitutive-reln}
\begin{aligned}
    T^{\mu\nu}_\text{1st} &= \lambda_1 (u^\mu B^\nu + u^\nu B^\mu) + \lambda_2 (u^\mu \omega^\nu + u^\nu \omega^\mu) \, , \\
    j^\mu_\text{1st} &= \zeta_1 B^\mu  + \zeta_2 \omega^\mu\,.
\end{aligned}
\end{equation}
These coefficients capture the flow of momentum and $\U$ current along the magnetic field line and vorticity.
By varying the effective action $W$ in \eqref{eq:genericW} with respect to the background metric and $\U$ gauge field, one can obtain expressions for $\zeta_i,\lambda_i$ in terms of the Chern--Simons coefficients $c_i$ appearing in the effective action \eqref{eq:genericW}~\cite{Banerjee:2012iz}, which we do not reproduce explicitly here.\footnote{Note that the 
direct variation of $W$ in \eqref{eq:genericW} results in the so-called \textit{consistent currents}, which are not invariant under a small gauge transformation $a\to a + d\lambda$. To obtain the currents of the form \eqref{eq:1stOrder-constitutive-reln}, one has to form \textit{covariant currents} by adding a Bardeen--Zumino term \cite{Bardeen:1984pm}. This procedure is well-documented for the equilibrium effective action setup in \S\S 2--3 of \cite{Banerjee:2012iz}. A modern review of this can also be found in \cite{Jensen:2013kka}. }

As suggested, the Chern--Simons coefficients $c_{0,1,2}$ in the effective action, and thus the transport coefficients $\lambda_{1,2}$ and $\zeta_{1,2}$, are closely related to anomalies in the underlying microscopic theory, which has long been appreciated in the case of perturbative anomalies (see {\em e.g.} \cite{Vilenkin:1980fu} and \cite{Landsteiner:2016led} for reviews). In the case of a theory with $\U$ and Poincar\'e global symmetry, the possible anomalies (which are strictly perturbative) are captured by the anomalous conservation law
\begin{equation}\label{eq:wardjcov}
    	\nabla_\mu j^\mu_{cov} = \frac{1}{4} \epsilon^{\mu \nu \rho \sigma} \left( 3 c_A f_{\mu \nu }f_{\rho \sigma} + c_m R^\alpha_{\;\; \beta \mu \nu} R^\beta_{\;\; \alpha \rho \sigma}  \right)\, .
\end{equation}
Here $c_A \propto \sum_i q_i^3$ and 
$c_m \propto \sum_i q_i$ are the perturbative anomaly coefficients, $j^\mu_{cov}$ is the \textit{covariant} $\U$ current,
and $f,R$ are the field strength and the Riemann curvature tensor associated to the $\U$ background gauge field and the metric, respectively. 
One can employ a macroscopic argument\footnote{One argument is to require that the entropy production for a theory with Ward identity \eqref{eq:wardjcov} and constitutive relation \eqref{eq:1stOrder-constitutive-reln} is positive definite \cite{Son2009,Neiman:2010zi}. See also \cite{Glorioso:2017lcn} where the same result is obtained using the KMS condition at the level of the effective action. Another argument, more in the spirit of direct anomaly matching, is to choose the coefficients $c_{0,1,2}$ to be appropriate functions of thermodynamic quantities ($T$ and $\mu$) such that the EFT in \eqref{eq:genericW} reproduces the Ward identity \eqref{eq:wardjcov}. \label{foot:pos-entropy}
} 
to show that the transport coefficients are almost entirely fixed in terms of the microscopic anomaly coefficients and the thermodynamic quantities $T$ and $\mu$, as follows:
\begin{equation}\label{eq:transport-coeff-perturbative-anom}
\begin{aligned}
\zeta_1 &= -6 c_A \mu\, ,\qquad & \zeta_2 &=  c_2 T^2 - 3 c_A \mu^2 \,, \\
\lambda_1 &=  c_2T^2-3c_A\mu^2\, , \qquad &\lambda_2 &=2 c_2 \mu T^2 -2 c_A \mu^3\,.
\end{aligned}
\end{equation}
At the moment, $c_2$ is a free coefficient of the mixed Chern--Simons term in (\ref{eq:genericW}).
These relations are consistent with the gradient expansion --  when one treats $\mu_B,T,g_{\mu\nu},a_\mu$ to be of order $\CO(\d^0)$, then the mixed $\U$-gravitational anomaly term multiplying $c_m$ in \eqref{eq:wardjcov} is of order $\CO(\d^4)$ while the rest are of order $\CO(\d^2)$.

In fact, one can go further: the response functions that encode the transport coefficients in \eqref{eq:transport-coeff-perturbative-anom} can be computed given a microscopic description. In both the perturbative r\'egime \cite{Ryu:2010ah,Landsteiner:2011cp,Loganayagam:2012pz,Golkar:2012kb} and in the large coupling r\'egime (via the holographic dual) \cite{Landsteiner:2011iq,Azeyanagi:2013xea,Grozdanov:2016ala}, it has been shown that the as-yet-unfixed coefficient $c_2$ is related to the gravitational anomaly coefficient $c_m$ via 
\begin{equation}\label{eq:ctilde-and-cm}
     c_2 = -8\pi^2 c_m\, .
\end{equation}
This strongly suggests there should be additional mechanisms that relate $c_2$ and $c_m$. Sure enough, the relation \eqref{eq:ctilde-and-cm} can also be derived using perturbative anomaly matching for the mixed anomaly between $\U$ and gravity,\footnote{An alternative derivation~\cite{Jensen:2012kj} for \eqref{eq:ctilde-and-cm} employs a particular derivation of the Cardy formula, which relates the thermodynamic pressure and the central charge by putting a theory on a cone~\cite{Bloete:1986qm}, to demand the consistency of the theory in the limit of small conical deficit angle to find the relation \eqref{eq:ctilde-and-cm} between $c_2$ and $c_m$. \label{foot:cone}
}
as in Refs.~\cite{Golkar:2015oxw,Chowdhury:2016cmh,Glorioso:2017lcn} -- see our discussion in footnote~\ref{foot:Golkar-Sethi}. In summary, the transport coefficients $\lambda_{1,2}$ and $\eta_{1,2}$ are fixed (in terms of $T$ and $\mu$) entirely by perturbative anomaly matching.

One might conclude from this that 
if all perturbative anomaly coefficients vanish, {\em i.e.} if $c_A = c_m =0$, then all the transport coefficients in 
\eqref{eq:1stOrder-constitutive-reln} should also vanish via \eqref{eq:transport-coeff-perturbative-anom}. In this paper we show that such a na\"ive statement need not be true when {\em discrete symmetries} in the underlying microscopic theory are taken into account. In particular, some of the transport coefficients in \eqref{eq:1stOrder-constitutive-reln} are compelled to be non-zero through a non-zero Chern--Simons coefficient $c_2$, when the additional $\mathbb{Z}_2$ symmetry has a mixed non-perturbative anomaly with the continuous $\U$ current. This effect is rather subtle; if one were to follow the standard hydrodynamics construction and list only the continuous global symmetries, then the role of the $\mathbb{Z}_2$ charge would be missed entirely. For instance, the `positivity of entropy production' argument (footnote~\ref{foot:pos-entropy}) cannot fix the coefficient $c_2$, because the Ward identities with and without an additional $\mathbb{Z}_2$ symmetry are identical. This is no surprise, since a non-perturbative anomaly, such as the original $\SU(2)$ anomaly of Witten~\cite{Witten:1982fp}, cannot be seen in Ward identities. Similarly, if one na\"ively applies the cone method of \cite{Jensen:2012kj} (footnote \ref{foot:cone}) while ignoring the possible $\mathbb{Z}_2$ holonomy, one would conclude that $ c_2 \propto c_m = 0$.

\section{A new mixed anomaly}\label{sec:formal-stuff}



\begin{table}[t]
  \centering
  \begin{tabular}{|c|ccccccc|}
    \hline
    $d$ & $~~~~~0~~~~~$ & $~~~~~1~~~~~$ & $~~~~~2~~~~~$ & $~~~~~3~~~~~$ & $~~~~~4~~~~~$ & $~~~~~5~~~~~$ & $~~~~~6~~~~~$\\
    \hline
    $\Omega^{\Spin}_d(B\U\times B\Z_2)$ & $\Z$ & $\Z_2^2$ & $\Z\times\Z_2^2$ & $\Z_2\times \Z_8$ & $\Z^2$ & $\Z_4$ & $\Z^2$\\
    \hline
  \end{tabular}
  \caption{The spin bordism groups for $B\U\times B\Z_2$ for degrees $0$ through $6$, which we compute in Appendix~\ref{sec:comp-with-ASS}. \label{fig:spin-bord-BU-BZ2-text}}
\end{table}

Following this hydrodynamical prelude, we now turn to the main subject of this paper.
Our mathematical starting point is the observation of a new non-perturbative anomaly in 4d quantum field theories with internal symmetry
    $G= \U \times \Z_2$
and defined with a spin structure, which means the microscopic theory has fermionic degrees of freedom. As reviewed in the Introduction, when perturbative anomalies cancel, the residual global anomaly is classified by the bordism group $\Omega_5^\Spin(BG)$, where $BG$ denotes the classifying space of the group $G$. In Appendix~\ref{sec:comp-with-ASS} we compute via the Adams spectral sequence that
\begin{equation} \label{eq:Z4}
    \Omega_5^{\mathrm{Spin}}(B\U \times B\Z_2) \cong \Z_4\, . 
\end{equation}
(We compute all bordism groups $\Omega^{\Spin}_d(B\U\times B\Z_2)$ in degrees $d=0$ through $6$, which we reproduce in Table \ref{fig:spin-bord-BU-BZ2-text} for reference.)
Eq. (\ref{eq:Z4}) tells us that the finest possible anomaly is $\Z_4$-valued. 
We moreover know that this must be a mixed anomaly, because neither $\U$ nor $\Z_2$ on its own carries a global anomaly in 4d, as follows from $\Omega_5^\Spin(B\U)=\Omega_5^\Spin(B\Z_2)=0$ \cite{Garcia-Etxebarria:2018ajm}. Indeed, the group $\Z_2$ on its own can have no chiral anomalies whatsoever simply by virtue of the fact that it has only {\em real} irreducible representations.
But by intertwining the discrete $\Z_2$ symmetry with a $\U$, we open up possibly anomalous complex representations, and a global anomaly can remain even when we cancel both perturbative anomalies associated with the $\U$ charge assignment.
We will show in this Section that this $\Z_4$ anomaly is carried by the matter content in Table~\ref{tab:min-content}.

Before we show this, it is helpful to first motivate the consideration of such a symmetry type, at the level of the microscopic QFT. It is not especially exotic or {\em ad hoc}; it can arise in otherwise run-of-the-mill weakly coupled theories.
For example, consider just a pair of left-handed Weyl fermions $\psi_{1,2}$ with a Yukawa coupling to a single real scalar $\Phi$ that is massive. In Weyl notation, where $\epsilon_{\alpha\beta}$ is used to contract the spinor indices, we have the Lagrangian
\begin{equation}
  \label{eq:toy-model}
    \mathcal{L}_{\mathrm{UV}} =
    i \psi_1^{\dagger}\bar{\sigma}^{\mu}\partial_{\mu}\psi_1 + i \psi_2^{\dagger}\bar{\sigma}^{\mu}\partial_{\mu}\psi_2 + \frac{1}{2}(\partial\Phi)^2 - \frac{1}{2}m^2 \Phi^2 + y \epsilon_{\alpha\beta}\psi_1^\alpha \Phi \psi_2^\beta + \text{h.c.}\, , \qquad y \in \mathbb{C}, \, m \in \mathbb{R}\, . 
\end{equation}
We want to know what are the global symmetries of this Lagrangian. A pair of 4d Weyl fermions on their own would, classically, have a pair of $\U$ symmetries, vector and axial, but these symmetries cannot be simultaneously preserved by the Yukawa coupling unless the real scalar $\Phi$ is charged.
But under any symmetry, the real scalar $\Phi$ can transform only up to a minus sign in order for the kinetic term $(\partial\Phi)^2$ to remain invariant. Therefore, there is a single $\U$ symmetry under which the fermion components are charged oppositely and $\Phi$ is neutral, which is the usual vector-like symmetry. The scalar kinetic and mass terms are however consistent with a $\Z_2$ symmetry that acts non-trivially on $\Phi$, provided exactly one of the Weyl components (say $\psi_1$, without loss of generality) is also charged, and the other Weyl is neutral. The global symmetry therefore acts as in Table~\ref{tab:min-content}.\footnote{Since the fermions have odd charges while the scalar's charge is even under $\U$, the most refined tangential structure for this toy model is, in fact, $\Spinc\times\Z_2$, where $\Spinc \cong (\Spin \times \U)/\Z_2$ where the quotient identifies the element $(-1)^F \in \Spin(4)$ with $e^{i\pi}\in \U$. Using this more refined tangential structure would allow us to define the theory on {\em any} orientable 4-manifolds, not just those that are spin. Nonetheless, the choice of either symmetry type will ultimately not affect the quantisation conditions we derive for the anomalous hydrodynamic transport -- see footnote~\ref{foot:spinc_MT}.}
\begin{table}[t]
\centering
\begin{tabular}{c|c|c}
        & $~~~\U~~~$ & $~~~\Z_2~~~$ \\
        \hline
        $~\psi_1~$ & $1$ & $1$\\
        $~\psi_2~$ & $-1$ & $0$\\
        \hline
        $~\Phi~$ & 0 & 1
\end{tabular}
\caption{Global symmetries of a pair of 4d Weyl fermions coupled to a massive real scalar. The fermion charges serve to generate the $\Z_4$-valued mixed anomaly at the heart of this paper. \label{tab:min-content}}
\end{table}

Indeed, at low energies $E < m$ 
one can integrate out the massive scalar $\Phi$ to obtain an EFT of only the Weyl fermions in Table~\ref{tab:min-content}, which are massless given that $\Phi$ does not condense. It is these massless fermions that are of interest to our discussion, because they carry any 't Hooft anomalies that obstruct gauging of the symmetry.

\subsection{Mapping torus construction}\label{sec:mapping-torus}

As mentioned, the bordism group computation (\ref{eq:Z4}) means that the finest phase one can obtain by computing $\exp(-2\pi i \eta)$ for any fermion representation of $G=\U \times \Z_2$ for which the perturbative anomalies vanish, and evaluated on any suitably spun mapping torus, is a fourth root of unity. Equivalently, four identical copies of any fermion representation must be anomaly-free. Our task is now to identify a fermion representation, and an explicit 5d mapping torus, for which the $\Z_4$-valued anomaly is exposed by computing the exponentiated $\eta$-invariant.

The 4d $\Z_4$ anomaly we are describing is in fact closely related to a well-known $\Z_8$-valued anomaly in 2d~\cite{Fidkowski_2010}, associated with a $\Z_2$-charged Majorana--Weyl fermion, which guides our construction of the mapping torus.\footnote{This 2d anomaly is in turn related to a $\Z_8$-valued parity anomaly in 1d, via the Smith isomorphism $\Omega_3^\Spin(B\Z_2) \cong \Omega_2^{\Pin^-} \cong \Z_8$~\cite{Tachikawa:2018njr,Hason:2020yqf}. }
Because we consider a fermion representation under the internal symmetry $G$ that is not real, which is a necessary condition for {\em any} chiral fermion anomaly in 4d, there is no analogue of the 2d Majorana--Weyl spinor available to us. Consequently, the bordism group cannot detect a $\Z_8$ phase: but, as we will see, the $\Z_4$ phase is closely related to the anomalous phase one would obtain for a {\em pair} of 2d Majorana--Weyls with $\Z_2$ charge, which together constitute a (complex) Weyl charged under $\U$.

The 5d mapping torus we consider, which turns out to be a valid representative of the generator of the bordism group $\Omega_5^\Spin(B\U \times B\Z_2)\cong \Z_4$, is a product manifold of the form
\begin{equation} \label{eq:mapping-torus-1}
M_5 = S^2 \times M_3,
\end{equation}
where $S^2$ has the $\U$ flux of a unit monopole through it, and $M_3$ is a particular
twisted 3-torus, which itself furnishes a generator of $\Omega^{\Spin}_3(B\Z_2)\cong \Z_8$, following the construction in
Ref.~\cite[\S 4]{Davighi:2020uab}. Let us recap that construction, which is important to the present discussion.

While we work in Euclidean signature, we keep in mind that one direction will be identified with time and hence inverse temperature when we pass to the hydrodynamic application. We denote this the $\tau$ direction, compactified via $\tau \sim \tau + \beta$. The $S^2$ factor in (\ref{eq:mapping-torus-1}) is in purely spatial directions. To construct $M_3$, start with a 3d cylinder of the form $S^1_\tau \times S^1_\theta \times I_x$, where $S^1_\tau$ is the circle in the (Euclideanised) time-direction, which we will refer to as a `thermal cycle', and $S^1_\theta$ is a circle in the remaining spatial direction of our $(3+1)$-d theory, parametrized by $\theta \sim \theta + L$. The remaining $I_x$ factor is, to begin with, an interval parametrized by an auxiliary coordinate $x \in [0,1]$ that takes us round the mapping torus (whose ends we are yet to glue). The spin structure corresponds to anti-periodic boundary conditions in both $\tau$ and $\theta$ directions. Finally, and crucially, we take the $\Z_2$ background gauge field to have non-trivial holonomy on 1-cycles that wrap around $S^1_\tau$, but trivial holonomy otherwise.

To form $X_3$, we then glue the ends of the cylinder but with a twist, by identifying
\begin{equation}
    (\tau, \theta, 0) \sim \left(\tau + \frac{2 \beta \theta}{L}, \theta, 1 \right)\, ,
\end{equation}
and taking the anti-periodic spin structure also around the new cycles parametrized by the (now-compactified) $x$-direction. We can think of this mapping torus as implementing a `large diffeomorphism', which the reader might recognize as (twice) the modular $\tT$-transformation on the $(\tau,\theta)$ torus, in the presence of both $\Z_2$ holonomy on the $\tau$ circle and $\U$ monopole flux through an auxiliary sphere. It is only with {\em both} $\Z_2$ holonomy {\em and} $\U$ flux that the mapping torus is not nullbordant, hence the conclusion that we are detecting a {\em mixed} anomaly.

The anomaly theory for chiral fermions is given by the exponentiated $\eta$-invariant~\cite{Witten:2019bou} of Atiyah, Patodi and Singer (APS)~\cite{Atiyah:1975jf,Atiyah:1976jg,Atiyah:1976qjr}, as reviewed in the Introduction. For the fermion representation of Table~\ref{tab:min-content}, perturbative gauge anomalies vanish ($\Phi_6=0$), meaning that $\exp(-2\pi i \eta)$ becomes a bordism invariant. So, by evaluating $\exp(-2\pi i \eta)$ on the mapping torus $M_5$ for this Dirac operator, we detect the bordism class $[M_5]\in \Omega_5^\Spin(B\U \times B\Z_2)$, as well as evaluating the anomalous phase accrued under the transformation described above.

\subsection{Computing the \texorpdfstring{$\eta$}{η}-invariant by anomaly interplay}

We can compute $\exp(-2\pi i\eta(M_5))$ using the idea of `anomaly interplay', as formulated in~\cite{Davighi:2020uab}, by which pushforwards and pullbacks are defined for bordism theories with different symmetry types $G$ and $G^\prime$ that are related by group homomorphisms. The specific calculation mirrors that of~\cite[\S 4]{Davighi:2020uab}, albeit in two dimensions higher and with an extra $\U$ symmetry factor.

The strategy is to embed the $\Z_2$ factor of $G$ inside a second $\U$ factor, call it $\U^\prime$. The mapping, which is a group homomorphism, is simply 
\begin{equation}
\pi:\Z_2 \to \U^\prime:\,(1 \text{~mod~} 2)\, \mapsto e^{i\pi} \in \U^\prime\, .
\end{equation}
Then, because
\begin{equation}
    \Omega_5^\Spin(B\U \times B\U^\prime)=0\, ,
\end{equation}
we know that $\pi_\ast M_5$ must be nullbordant in $\Omega_5^\Spin(B\U^2)$, meaning that the mapping torus can be realised as the boundary of a 6-manifold by embedding the $\Z_2$ connection inside a $\U^\prime$ connection. As in~\cite[\S 4]{Davighi:2020uab}, an explicit $\U^\prime$ connection on $X_3$ that matches the desired $\Z_2$ holonomies is given by $A^\prime(\tau,\theta,x) = \frac{1}{2}d\tau$. This can be extended to a 4-manifold $Y_4$ by filling in a disc $D^2$ bounded by $S^1_\tau$ with radial coordinates $(r\in [0,1],\tau)$, such that $\partial Y_4 = X_3$, via $A^\prime(r,\tau,\theta,x)=\frac{1}{2}r d\tau+ (1-r) x d\theta$, which one can check is compatible with the mapping torus gluing relations. For this $A^\prime$, we integrate~\cite{Davighi:2020uab}
\begin{equation}
    \frac{1}{8\pi^2}\int_{Y_4} f^\prime \wedge f^\prime = \frac{1}{4}\, ,
\end{equation}
where $f^\prime = dA^\prime$.

Having realised the (pushed-forward) mapping torus $\pi_\ast M_5$ as the boundary of a 6-manifold $M_6 = S^2 \times Y_4$, the APS index theorem~\cite{Atiyah:1975jf,Atiyah:1976jg,Atiyah:1976qjr} allows us to evaluate the exponentiated $\eta$-invariant simply by integrating the differential form $\Phi_6$:
\begin{equation} \label{eq:APS}
    \exp \left( -2\pi\ii \eta(M_5) \right) = \exp(-2\pi\ii \int_{M_6}\Phi_6)\, .
\end{equation}
Here, $\Phi_6$ is the anomaly polynomial for the particular Dirac operator under consideration:
\begin{equation} \label{eq:Phi6}
    \Phi_6 = \hat{A}(R) \Tr \exp\left(\frac{F}{2\pi} \right)\bigg|_{6} = -\frac{1}{48 \pi}p_1(R) \Tr F + \frac{1}{8\pi^3} \Tr F^3\, ,
\end{equation}
where $\hat{A}(R)$ and $p_1(R)$ denote the Dirac genus and first Pontryagin number of the tangent bundle of $M_6$.
For our pair of $\U$ background fields, we can write $F=f t + f^\prime t^\prime$, where $t^{(\prime)}$ denotes the single Lie algebra generator of $\U^{\prime}$. 

Considering a generic spectrum of 4d left-handed Weyl fermions $\psi_i$, $i=1,\ldots,n$, with charges $(Q_i\mod 2,q_i)$ under $G= \Z_2 \times \U$, such that $\sum_iq_i^3 = \sum_i q_i =0$ to ensure the cancellation of perturbative anomalies, the embedding $\pi:\Z_2 \subset \U^\prime$ is consistent with `pushed forward' charges
$\tilde{Q}_i$ that satisfy $\tilde{Q}_i \cong Q_i\mod 2$. The $\U$ generators $t,t^{\prime}$, which are just charge matrices, are now explicitly given by $t = \diag \left( q_1,\ldots, q_n \right)$, $t^{\prime}=\diag ( \tilde{Q}_1,\ldots, \tilde{Q}_n)$.
When integrated on $M_6$, only the cross-term $\sim f f^{\prime 2}$ in $\Phi_6$ survives. Using (\ref{eq:APS}) and (\ref{eq:Phi6}), we get
\begin{equation}
\exp(-2\pi\ii \eta(M_5))  = \exp \left(  -\pi\ii \mathcal{C} \int_{S^2} \frac{f}{2\pi} \int_{Y_4} \frac{f^\prime \wedge f^\prime}{(2\pi)^2}  \right)= \exp(-\frac{\pi\ii \mathcal{C}}{2}) \, ,
\end{equation}
where we define the anomaly coefficient $\mathcal{C}$ to be
\begin{equation}
    \mathcal{C} := \sum_i q_iQ_i^2 \in \Z\, .
\end{equation}
The anomalous phase on this mapping torus is indeed an element of $\Z_4$, with the anomaly given by
$\mathcal{C}\mod 4$. As a consistency check, we note that this anomaly formula depends only
on the original $\Z/2$ charges $Q_i$ and not on the choice of lift to $\U^\prime$ charges, because it is well-defined
under $\tilde{Q}_i \to \tilde{Q}_i+2r_i$ for any integers $r_i$.

The minimum value $\mathcal{C}=1\mod 4$ is realised by the fermion content in Table~\ref{tab:min-content}. This computation also establishes that $[M_5] = 1\mod 4\in \Omega_5^\Spin(B\U \times B\Z_2) \cong \Z_4$, and so the mapping torus we constructed does indeed provide a generator for the bordism group in question.



\section{Fractional transport from the discrete anomaly}\label{sec:transport}
In this Section, we show how this non-perturbative mixed anomaly is manifest in the hydrodynamic limit of a theory with $\U\times \Z_2$ symmetry type, by using the
mapping torus $M_5$ constructed previously to constrain the equilibrium effective action discussed in \S \ref{sec:hydro-review}. In \S \ref{sec:EFT-constraint} we show how to adapt the anomaly-matching method first employed in Ref.~\cite{Golkar:2015oxw}, there for a theory of Weyl fermions with (perturbative) mixed $\U$-gravitational anomaly (see footnote~\ref{foot:Golkar-Sethi}), to the system with $G = \U\times \mathbb{Z}_2$ global symmetry. 
The mapping torus $M_5$ encodes how the partition function, evaluated in a particular background field configuration, transforms under a large diffeomorphism; by requiring that the hydrodynamic effective action (\ref{eq:genericW}) reproduces the same variation, one can relate the coefficients $c_i$ of the effective action, and consequently the transport coefficients $\lambda_i$ and $\zeta_i$, to the non-perturbative anomaly in the microscopic theory.

The mapping torus method's prediction for the transport coefficients should also be confirmed in a microscopic theory that realised the mixed $\U\times \mathbb{Z}_2$ anomaly, in a r\'egime in which the transport coefficients are calculable. We perform such a non-trivial check in \S \ref{sec:free-field-transport}, in a microscopic theory with the fermion content in Table~\ref{tab:min-content}. Taking a limit in which the interactions decouple,
we reproduce the predicted values of the transport coefficients obtained in \S \ref{sec:EFT-constraint} by the general anomaly matching argument. 

\subsection{Anomaly matching constraints on the hydrodynamic effective action}\label{sec:EFT-constraint}


The goal is to arrive at the effective action $W = -\log Z$, where $Z$ is the thermal partition function of a theory with $\U\times \mathbb{Z}_2$ global symmetry on a four dimensional spin-manifold $X_4$. As in \S \ref{sec:hydro-review}, $X_4$ is a thermal-cycle-fibration over a spatial 3-manifold $X_3$, {\em viz.} $S^1_\tau \hookrightarrow X_4 \to X_3$. The partition function is 
\begin{equation}\label{eq:partition-func-def}
    Z = \tr\left[\exp\left( -\beta (\hat H - \mu \hat q) + i w_1 \hat Q   \right)  \right]\, ,
\end{equation}
where $\hat q$ and $\hat Q$ are the charge operators for the $\U$ and $\mathbb{Z}_2$ symmetries, respectively, defined on the spatial slice $X_3$. 
The term $\beta \mu$ is the analytic continuation of the $\U$ holonomy around the thermal cycle, and $w_1$ is the $\mathbb{Z}_2$ holonomy which takes value in $H^1(X_4;\mathbb{Z}_2)$. When $w_1=1\text{~mod~} 2$, the term $e^{iw_1\hat{Q}}$ introduces an insertion of the operator $e^{i\hat Q}$ on the thermal circle, which flips the boundary condition from periodic to anti-periodic and {\em vice versa} for the fields charged under the $\mathbb{Z}_2$ symmetry. The most general form of $W=-\log Z$ is the same as in \eqref{eq:genericW}. 

For this partition function to describe a theory with the mixed anomaly described in the previous Section, it must transform appropriately under the modular-$\mathtt{T}$ transformation described in \S \ref{sec:mapping-torus} above, when placed on a background $X_4 = S^2\times S^1_\tau\times S^1_\theta$ with $\mathbb{Z}_2$ holonomy $w_1\ne 0$ around the thermal cycle $S^1_\tau$ and the background $\U$ monopole $\int_{S^2} f\ne 0$. More precisely, the partition function must transform as 
\begin{equation}\label{eq:ZvsZT2}
  Z \to   Z^{\mathtt{T}^2} = Z \exp\left( -2\pi i \eta(M_5)\right)
\end{equation}
where $\mathtt{T}^2$ is a modular transformation on the torus $S^1_\tau \times S^1_\theta$ which takes 
\begin{equation}\label{eq:T2-transf}
\mathtt{T}^2: \tau \to \tau +\frac{2 \beta}{L}\theta\,,\qquad \theta \to \theta\, ,    
\end{equation}
with $L$ being the circumference of $S^1_\theta$, and $\eta(M_5)$ being the APS $\eta$-invariant evaluated on the mapping torus as discussed in detail in~\S \ref{sec:mapping-torus}. 

The hydrodynamic equilibrium partition function \eqref{eq:genericW} can capture the variation \eqref{eq:ZvsZT2}. We follow a similar argument to that proposed by Golkar and Sethi ~\cite{Golkar:2015oxw}.
First, notice that under the large diffeomorphism transformation \eqref{eq:T2-transf} the KK gauge field is transformed, per \eqref{eq:time-shift}, as 
\begin{equation}\label{eq:alpha-under-T2}
\mathtt{T}^2: \alpha_\theta \to \alpha_\theta + \frac{2\beta}{L}    \, ,
\end{equation}
with the $\U$ gauge field $\CA_i$, the time-component $a_t$, and other components left unchanged. The effective action in \eqref{eq:genericW} thus transforms as 
\begin{equation}
    W \to W^{\mathtt{T}^2} = W +2i c_2 \int_{S^2} d\CA \,.
\end{equation}
Thus, for the effective action $W=-\log Z$ to match the anomalous variation \eqref{eq:ZvsZT2} in the fundamental theory with mixed $\U\times \mathbb{Z}_2$ anomaly,
the hydrodynamic effective description  must satisfy 
\begin{equation} \label{eq:M5-anomaly-matching}
    -i(W^{\mathtt{T}^2} - W) = 2c_2 \int_{S^2} d\CA = 2\pi \eta(M_5)\, .
\end{equation}
For the mapping torus $M_5$ with unit $\U$ monopole flux through the spatial $S^2$ factor, we have $\int d\CA = 2\pi$, and thus\footnote{
We do not expect the answer to change when the more refined symmetry type $\Spinc \times \Z_2$ of the toy model Eq.~\eqref{eq:toy-model} is used instead, although the anomalies are now classified by $\Omega^{\Spinc}_5(B\Z_2) \cong \Z_8 \times \Z_2$ \cite{bahri1987eta} (which are related to reflection anomalies in 3d \cite{Chen:2023hmm}). Heuristically, the finest mod 8 anomaly should be detectable by a non-spin orientable manifold that generates the $\Z_8$ factor in the bordism group, allowing us to take $\int d\CA/2\pi$ to be half-integers. But the extra factor of $1/2$ is cancelled in Eq. (\ref{eq:M5-anomaly-matching}) by the same extra factor in the $\eta$-invariant, which is now $1/8$ rather than $1/4$. The upshot is that the quantisation condition on $c_2$ is unaffected.
\label{foot:spinc_MT}
}
\begin{equation}
c_2 = \frac{1}{2} \eta[M_5] = \begin{cases}
    \frac{1}{2} \left( \frac{1}{4} \text{ mod }1 \right) & \text{ if } w_1 =1\,,\\
    0 &\text{ if } w_1 = 0\, .
\end{cases}
\end{equation}
One might ask whether or not this analysis really fixes the coefficients in the effective action; in particular, are there constraints on the other Chern--Simons coefficients $c_0$ and $c_1$? Firstly, all the Chern--Simons coefficients $c_{0,1,2}$ cannot depend on the thermodynamic variables $T$ or $\mu$ when the system has no perturbative anomaly (which means that $W$ has to be invariant under small gauge transformations and diffeomorphisms). If the $c_{0,1,2}$ are integer-quantised, then CPT symmetry enforces both $c_0$ and $c_1$ to vanish as pointed out in \cite{Banerjee:2012iz}, and as we reviewed in \S \ref{sec:hydro-review}. The remaining possibility is that $c_0$ and $c_1$ are non-zero fractional Chern--Simons couplings. However, the fact that $\Omega^\text{Spin}_5(B\U\times B\mathbb{Z}_2) \cong \mathbb{Z}_4$ is already realised by the mapping torus $M_5$, plus the fact that the bordism group $\Omega^\text{Spin}_5(B\U)$ vanishes, suggests that there is no background configuration and large diffeomorphism that requires $c_0$ or $c_1$ to be non-zero. 

With all coefficients $c_i$ in the effective description fixed, one can now derive the consequences for the hydrodynamic transport coefficients $\lambda_i$ and $\zeta_i$, as defined in \eqref{eq:1stOrder-constitutive-reln}.
We obtain the constitutive relation in \eqref{eq:1stOrder-constitutive-reln} with the {\em fractionally quantised} transport coefficients:  
\begin{equation}\label{eq:transport-from-anomaly}
    \zeta_1 = 0 \, , \qquad \zeta_2 = \lambda_1 = -\frac{T^2}{2} \left(\frac{1}{4}\text{ mod }1  \right)\, , \qquad \lambda_2 = \frac{\mu T^2}{2}\left(\frac{1}{4} \text{ mod }1  \right)\, ,
\end{equation}
which will be further verified by a direct computation in the following Subsection.
It should also be emphasised that the $\mathbb{Z}_2$ holonomy plays a crucial role in this effect, as all the above transport coefficients vanish without it. 
Physically, one can think of the $\Z_2$ holonomy as providing a choice of valid boundary conditions for quantising our $\Z_2$-charged microscopic fields.

While the prediction of non-trivial hydrodynamic transport coefficients due to `mapping torus constraints' has been considered in \cite{Golkar:2015oxw} (see also \cite{Chowdhury:2016cmh,Glorioso:2017lcn}), we wish to emphasise two important differences between the existing literature and the example presented here. First, the aforementioned works consider theories with perturbative anomalies (in particular, perturbative mixed anomalies between $\U$ and gravity) as we discussed in footnote~\ref{foot:Golkar-Sethi}, while the anomaly considered in this work is a genuine non-perturbative anomaly which does not violate the conservation of the energy-momentum tensor or $\U$ current at any level in the derivative expansions. Second, the global $\mathbb{Z}_2$ symmetry, which plays a crucial role in this anomaly matching, is often ignored in the study of transport and hydrodynamics as it does not have a Noether current to enter systems of differential equations. We argue that neglecting such discrete global symmetries may not always be completely justified; non-trivial discrete holonomy can still leave its mark on the continuous $\U$ current through the mixed $\U\times \mathbb{Z}_2$ anomaly. 

\subsection{Explicit computations in the free fermion limit}\label{sec:free-field-transport}

It is well-known in linear response theory that the transport coefficients can be extracted from certain limits of the retarded 2-point correlation functions of the Noether currents via the Kubo formalism (see {\em e.g.} \cite{Kadanoff:1963,Kovtun:2012rj} for reviews). The associated {\em Kubo formulae}, namely appropriate 2-point functions and limits, for the transport coefficients in this work can be derived directly from the constitutive relation \eqref{eq:1stOrder-constitutive-reln}. They are found to be \cite{Landsteiner:2012kd,Chowdhury:2015pba}\footnote{Note that the formulae in \eqref{eq:Kubo-formulae} are slightly different from those presented in the mentioned references, where the transport coefficients are written schematically as 
\begin{equation}
 \lim_{k\to 0}\lim_{\omega\to 0} \frac{1}{ik} \la \CO_A(\omega,k)\CO_B(-\omega,-k) \ra_R\,.   \nonumber
\end{equation}
This derivation is equivalent to ours provided that (i) $\text{Re}\la \CO_A \CO_B\ra = \CO(k^2)$ and (ii) $\text{Im}\la\CO_A \CO_B\ra \propto k + \CO(k^2)$ in the limit of zero frequency and large wavelength. The requirement (i), however, may not hold due to the contact term. Hence, we adopt the alternative prescription \eqref{eq:Kubo-formulae} for the Kubo formulae to avoid this unnecessary requirement.
}
%
%
\begin{equation}\label{eq:Kubo-formulae}
\begin{aligned}
    \zeta_1 &= \frac{1}{2}\lim_{k_m \to 0}\lim_{\omega \to 0} \epsilon_{ijm}\frac{\d }{\d k_m} \text{Im } \la j^i(\omega,k_m) j^j(-\omega,-k_m)\ra_R \,,\\
    \zeta_2 &= \frac{1}{2}\lim_{k_m \to 0}\lim_{\omega \to 0} \epsilon_{ijm}\frac{\d }{\d k_m} \text{Im } \la j^i(\omega,k_m) T^{0j}(-\omega,-k_m)\ra_R \,,\\
    \lambda_1 &= \frac{1}{2}\lim_{k_m \to 0}\lim_{\omega \to 0} \epsilon_{ijm}\frac{\d }{\d k_m} \text{Im } \la T^{0i}(\omega,k_m) j^j(-\omega,-k_m)\ra_R \,,\\
    \lambda_2 &= \frac{1}{2}\lim_{k_m \to 0}\lim_{\omega \to 0} \epsilon_{ijm}\frac{\d }{\d k_m} \text{Im } \la T^{0i}(\omega,k_m) T^{0j}(-\omega,-k_m)\ra_R 
\end{aligned}
\end{equation}
The subscript $R$ denotes retarded 2-point correlation functions, obtained via 
\begin{equation}
    \la \CO_A(\omega,k) \CO_B(-\omega,-k)\ra_R = \frac{1}{\sqrt{-g}} \frac{\delta^2\, \log Z}{\delta \phi_A(-\omega,-k)\delta \phi_B(\omega,k)}\Big\vert_{\phi_A,\phi_B =0}\,\, ,
\end{equation}
where $\phi_A = \{ \delta g_{\mu\nu}, \delta a_\mu\}$ being the source of the operator $\CO_A = \{ T^{\mu\nu}, j^\mu\}$.
The correlation functions listed in \eqref{eq:Kubo-formulae} are static correlation functions and are accessible from the equilibrium effective description in \eqref{eq:genericW}.

In this Subsection, we use these Kubo formulae to calculate the transport coefficients $\lambda_{1,2}$ and $\zeta_{1,2}$ given an explicit microscopic theory that carries our 1 mod 4 mixed anomaly in $\U \times \Z_2$ symmetry. We consider the fermion content written in Table~\ref{tab:min-content}, in the limit of zero interactions.
Technically-speaking, the free theory presented here does not have a well-defined hydrodynamic limit whereby the length- and time-scales of the collective excitation are larger than the microscopic interaction scale (which, formally, is infinite). But crucially, if these transport properties are controlled by the global anomaly, then they are rigidly quantised in a way that cannot vary smoothly in the coupling constants. Thus, the transport coefficients obtained via the free field computation should be in agreement with those from hydrodynamic consistency conditions in \eqref{eq:transport-from-anomaly}. 

\begin{figure}[h!]
  \centering
  \begin{subfigure}[b]{0.45\textwidth}
    \centering
    \includegraphics[scale=1]{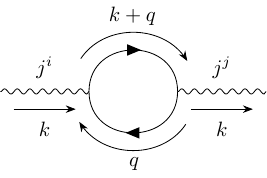}
    \caption{\label{fig:JJ-feyn}}
  \end{subfigure}
    \begin{subfigure}[b]{0.45\textwidth}
     \centering
     \includegraphics[scale=1]{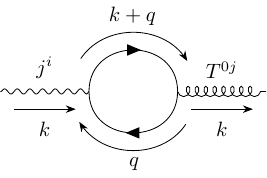}
     \caption{\label{fig:JT-feyn}}
  \end{subfigure}
  \caption{The relevant Feynman diagrams for (a) the $j$-$j$ and (b) the $j$-$T$ correlators. These diagrams and those in Fig.~\ref{fig:feynman-diag-TT} are drawn with the Ti{\it k}Z-Feynman package \cite{Ellis:2016jkw}.}
  \label{fig:feynman-diag}
\end{figure}

\begin{figure}[h!]
  \centering
  \begin{subfigure}[b]{0.45\textwidth}
    \centering
    \includegraphics[scale=1]{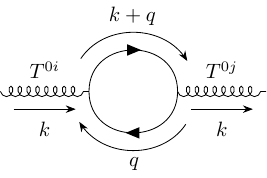}
    \caption{\label{fig:TT-feyn-bubble}}
  \end{subfigure}
    \begin{subfigure}[b]{0.45\textwidth}
     \centering
     \includegraphics[scale=1]{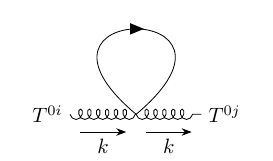}
     \caption{\label{fig:TT-feyn-seagull}}
  \end{subfigure}
  \caption{The relevant Feynman diagrams that contribute to the $T$-$T$ correlator: (a) the bubble diagram, and (b) the seagull diagram.}
  \label{fig:feynman-diag-TT}
\end{figure}

Similar computations of transport coefficients can be found in Refs.~\cite{Kharzeev:2009pj,Landsteiner:2011cp,Manes:2012hf,Landsteiner:2012kd,Chowdhury:2015pba}. They boil down to computing the 1-loop Feynman diagrams shown in Fig.~\ref{fig:feynman-diag} and Fig.~\ref{fig:feynman-diag-TT} in the Euclidean theory, which involves performing the Matsubara sum over frequencies, and then analytically continuing to obtain the retarded 2-point functions that appear in the Kubo formulae. For a single Weyl fermion with anti-periodic boundary condition around the thermal cycle $S^1_\tau$, one finds 
\begin{equation}\label{eq:transport-coeff-SingleWeyl}
\begin{aligned}
\zeta_1^{single}&= \frac{1}{4\pi^2} \int^\infty_0 dq f(q,\mu)\,,\\
\zeta_2^{single} = \lambda_1^{single}&= \frac{1}{4\pi^2} \int^\infty_0 dq \, q \bar f(q,\mu)\,,\\
\lambda_2^{single} &= \frac{1}{8\pi^2} \int^\infty_0 dq\, q^2 f(q,\mu)\, ,
\end{aligned}
\end{equation} 
where $f$ and $\bar f$ are combinations of the Fermi--Dirac distribution $n_F(q,\mu) = \left( e^{\beta(q-\mu))}+1 \right)^{-1}$, namely
\begin{equation}
    f(q,\mu) = n_F(q,\mu)- n_F(q,-\mu) \, , \quad \bar f(q,\mu) = n_F(q,\mu) + n_F(q,-\mu)\, .
\end{equation}
The integrals in Eq.~\eqref{eq:transport-coeff-SingleWeyl} can be evaluated and expanded in the limit $T\gg \mu$, to obtain all the transport coefficients for a theory with $\U^3$ and mixed $\U$-gravitational anomalies, as in \eqref{eq:transport-coeff-perturbative-anom}.



The above computation of the transport coefficients can easily be adapted to the free fermion theory with mixed $\U\times \mathbb{Z}_2$ anomaly in \S \ref{sec:formal-stuff}. The key idea is to realise that, upon turning on the $\mathbb{Z}_2$ holonomy, the symmetry operator $e^{ix_1 \hat q}$ in \eqref{eq:partition-func-def} is activated and the operator $\hat q$ flips the boundary condition of $\psi_1$, which is charged under $\mathbb{Z}_2$, while the boundary condition of $\psi_2$, neutral under $\mathbb{Z}_2$, remains anti-periodic. Upon turning on the chemical potential for the $\U$ symmetry, we find that
\begin{equation}
    \psi_1(\tau+\beta) = e^{\beta\mu}\psi_1(\tau) \, , \qquad \psi_2(\tau+\beta) = -e^{-\beta\mu} \psi_2(\tau)
\end{equation}
where the chemical potential $\mu$ appears in the exponentials with opposite sign for $\psi_1$ and $\psi_2$, due to their opposite $\U$ charge assignment as in Table \ref{tab:min-content}. Due to these different boundary conditions, the Euclidean propagators for fermions $\psi_s$, with $s=1,2$, are  
\begin{equation}\label{eq:freepropagator}
	S_s(q) = \frac{1}{2}\sum_{t=\pm} \Delta_t(i\omega^{(s)}_n + (-1)^{s+1} \mu,q  )P_+ \gamma_\mu \hat q^\mu_t\, ,\qquad \Delta_t(ix+y, q) = \frac{1}{ix + y  -t E_q}\,,
\end{equation}
where we follow the notation in {\em e.g.} \cite{Kharzeev:2009pj}, and where $P_+ = \frac{1}{2}(1+\gamma_5)$ is the usual projection operator.
Hatted vectors are defined to be $\hat q_t = (1, t q^i/E_q)$ with $E_q = |{\bf q}|$. The set of Matsubara frequencies $\omega^{(s)}_n$ for each fermion mode reflects the boundary conditions, namely
\begin{equation}
	\omega^{(1)}_n = \frac{2\pi  n}{\beta} \, , \qquad \omega^{(2)}_n = \frac{\pi}{\beta} (2n+1)\, , 
\end{equation}
where $n \in \mathbb{Z}$.

Plugging these propagators into the Feynman diagrams in Figs.~\ref{fig:feynman-diag} and \ref{fig:feynman-diag-TT}, one can find the contributions to the transport coefficients from each fermion species.
Those due to $\psi_2$ are the same as the contributions \eqref{eq:transport-coeff-SingleWeyl} from a single Weyl fermion (modulo the sign of the chemical potential), while those from $\psi_1$ have a slightly different form due to the difference at the stage of the Matsubara frequency sum. To see this, let us first introduce the functions
\begin{equation}
    b(q,\mu) = n_B(q,\mu) - n_B(q,-\mu) \, , \qquad \bar b(q,\mu) = n_B(q,\mu) + n_B(q,-\mu)\,,
\end{equation}
where $n_B(q,\mu) = \left(e^{\beta(q-\mu)}-1\right)^{-1}$ is the Bose--Einstein distribution function. Then,  combining the contributions from both fermions, we find the resulting transport coefficients to be 
\begin{equation}\label{eq:result-transport-coeff}
    \begin{aligned}
\zeta_1 &= \frac{1}{4\pi^2} \text{Re}\int^\infty_0 dq \left( f(q,\mu)+ b(q,\mu) \right) = 0\,,\\
\zeta_2 &= -\frac{1}{4\pi^2} \text{Re}\int^\infty_0 dq \,q\left( \bar f(q,\mu) + \bar b(q,\mu) \right)\\ 
&= \frac{T^2}{8\pi^2} \text{Re}\left[\mathrm{Li}_2(e^{\frac{2\mu}{T}})+\mathrm{Li}_2(e^{-\frac{2\mu}{T}})-4\mathrm{Li}_2(e^{\frac{\mu}{T}})-4\mathrm{Li}_2(e^{-\frac{\mu}{T}})\right]\\
&= -\frac{1}{2\pi^2} \left( \frac{\pi^2T^2}{4} \right)\,,
\end{aligned}
\end{equation}
where $\mathrm{Li}_2(z)=-\int_0^z \frac{\log(1-t)}{t} dt$ is the di-logarithm function,
and
\begin{equation}
\label{eq:result-transport-coeff-2}
\begin{aligned}
\lambda_1 &=\zeta_2\,,\\
\lambda_2 &= \frac{1}{8\pi^2} \text{Re}\int^\infty_0 dq \, q^2\left( f(q,\mu) + b(q,\mu) \right)\\
&= \frac{1}{8\pi^2} \frac{T^3}{2}\text{Re}\left[\mathrm{Li}_3(e^{-\frac{2\mu}{T}})-\mathrm{Li}_3(e^{\frac{2\mu}{T}}) +8\mathrm{Li}_3(e^{\frac{\mu}{T}})-8\mathrm{Li}_3(e^{-\frac{\mu}{T}})\right] \\
&= \frac{1}{8\pi^2} \left( \pi^2\mu T^2 \right)\,,
    \end{aligned}
\end{equation}
where $\mathrm{Li}_3(z)=\int_0^z \frac{\mathrm{Li_2(t)}}{t} dt$ is the tri-logarithm function. Note that the final results in Eqs. \eqref{eq:result-transport-coeff} and \eqref{eq:result-transport-coeff-2} are exact, due to the inversion formulae for polylogarithms.
Had one used the anti-periodic boundary condition for $\psi_1$, corresponding to the case where the discrete $\mathbb{Z}_2$ holonomy is turned off, the computation can be carried out in the same manner. In that case, one finds that the resulting transport coefficients take the same form but with $b(q,\mu)$ replaced by $-f(q,\mu)$. In this case, all the transport coefficients vanish.

The transport coefficients obtained in \eqref{eq:result-transport-coeff} and \eqref{eq:result-transport-coeff-2} are in perfect agreement with those we previously obtained from the effective action approach in \S \ref{sec:EFT-constraint}, see Eq.~\eqref{eq:transport-from-anomaly}. We again see that the non-trivial $\Z_2$ holonomy $w_1 \in H^1(M_4;\mathbb{Z}_2)$ plays a crucial role, as we did before from our $\eta$-invariant computation; from the perspective of the present Subsection, this is seen by the flipping of the boundary condition for one of the two fermion modes, that is necessary to obtain non-zero transport coefficients. 

\section{Discussion and outlook}\label{sec:conclusion}

The purpose of this work is twofold. Firstly, we uncover a new $\mathbb{Z}_4$-valued non-perturbative anomaly in 4d quantum field theories with $G=\U \times \mathbb{Z}_2$ global symmetry (and spin structure), that can be realised in a very simple system. 
Following now-standard lore, if a 4d QFT with global symmetry $G$ has no perturbative anomaly, then the exponentiated APS $\eta$-invariant which controls the transformation properties of the fermion partition function becomes a bordism invariant, living in the group $\hom \left(\mathrm{Tor}\, \Omega_{5}^\Spin(BG), \U\right)$. We compute the bordism group $\Omega^\text{Spin}_5(BG)$ to be the cyclic group $\Z_4$ using the state-of-the-art Adams spectral sequence.
We explicitly construct a 5d mapping torus that serves as a generator of the bordism group, and employ
`anomaly interplay' to not only confirm that the $\eta$-invariant on the mapping torus is indeed $\mathbb{Z}_4$-valued, but also to determine the charge assignments in the microscopic theory that exhibit this new anomaly. 

Our second main purpose is to study the effect of this new $\U\times \mathbb{Z}_2$ anomaly at the level of transport phenomena in the hydrodynamic effective description of a theory at finite temperature and $\U$ chemical potential. We find that the anomaly gives rise to a (non-dissipative) current and momentum flow along the magnetic field and/or vorticity similar to the chiral magnetic and chiral vortical effects, despite the absence of any perturbative anomaly. 
We derive these conditions from two viewpoints: (i) by demanding consistency of the hydrodynamic constitutive relation and the thermal partition function with the microscopic anomaly, and (ii) via a direct EFT matching computation starting from a microscopic theory in the free-fermion limit; the two calculations are in perfect agreement. In doing so, we also hope to further clarify how to utilise information from the bordism group and its generator to constrain unknown coefficients in the equilibrium effective action which, in turn, determines the anomaly induced transport coefficients.

The essential role of the $\mathbb{Z}_2$ symmetry, in particular of the non-trivial $\mathbb{Z}_2$ holonomy, should be emphasised -- without the $\Z_2$, a continuous $\U$ current on its own cannot exhibit any global anomaly. A non-trivial $\Z_2$ holonomy is crucial in allowing the exponentiated $\eta$-invariant to yield a non-trivial phase, which results in the non-trivial anomalous transport phenomena; from the viewpoint of the free-fermion computation, the $\Z_2$ holonomy effectively switches a fermionic boundary condition in the microscopic computation. A discrete global symmetry such as this, which goes beyond C, P and T symmetry, can be subtle and is often ignored in the construction of low-energy effective theories. This is especially true for a conventional construction of hydrodynamics as  a system of differential equations involving the Noether currents. There, the discrete global symmetry would be completely discarded, and the effect of the $\U\times \mathbb{Z}_2$ anomaly missed entirely. The anomaly we study in this paper demonstrates that \textit{all} global symmetries, not just the continuous ones, can be important in the hydrodynamic r\'egime. It would therefore be interesting to revisit the said limit of theories with known discrete and continuous symmetries for which the appropriate bordism groups are non-trivial, to identify other examples of fractional anomalous transport induced by non-perturbative anomalies. 

The existence of the new mod 4 anomaly opens up a number of interesting future directions. An immediate follow-up question is what happens to the anomaly when we condense the real scalar field that couples to the Dirac fermion in the toy model introduced in \S \ref{sec:formal-stuff}. In the condensed phase, the $\Z_2$ breaks spontaneously and there exist domain walls separating different vacua. The $\Z_2$ unitary global symmetry descends to an anti-unitary time-reversal symmetry T with T$^2=1$, giving rise to the tangential structure $\Pin^-\times \U$ on the domain wall~\cite{Hason:2020yqf, Debray:2023ior}. We would like to determine explicitly how the $\Z_4$-valued anomaly in 4d is accounted for in the domain wall theory. The anomalies of the latter theory are classified by the bordism group $\Omega^{\Pin^-}_4(B\U)$  which, sure enough, is also isomorphic to $\Z_4$ as we show in Appendix~\ref{sec:Pin-bord-BU1}. 

Another area worthy of further exploration is whether this anomaly can provide us with new examples of {\em symmetric mass generation,} a phenomenon whereby fermions become massive without breaking a chiral symmetry through a strongly coupled interaction. A well-known example occurs in the Fidkowski--Kitaev spin chain model \cite{Fidkowski_2010}, which is in fact directly related at the bordism level to the mapping torus we construct in \S~\ref{sec:mapping-torus} of this paper. 
It is widely believed that symmetric mass generation can be realised whenever the chiral symmetry is anomaly-free. Thus, in principle we should be able to trivially gap out a system of 4 Dirac fermions without breaking its $\U \times \Z_2$ chiral symmetry, even though the naïve mass term does not allow it. One mechanism for symmetric mass generation that has been explored in detail {\em e.g.} in Refs.~\cite{Razamat:2020kyf,Tong:2021phe}, that one could imagine might be used to gap our 4d Dirac fermions,
is through the phenomenon of {\em s-confinement}~\cite{Intriligator:1995id}, whereby confinement occurs without chiral symmetry breaking in $\mathcal{N}=1$ supersymmetric gauge theories. We aim to explore this possibility in future work.


A holographic realisation of this mixed anomaly would also be an interesting direction of study. In contrast to the perturbative anomaly, where the bulk anomaly theory is provided by the Chern--Simons differential form, the $\mathbb{Z}_4$ torsion nature of the mixed $\U\times \mathbb{Z}_2$ anomaly implies that its corresponding bulk anomaly theory cannot be written in terms of a differential form, and that a more general mathematical structure such as those in \cite{Hsieh:2020jpj,Apruzzi:2021nmk} is required. Finding a holographic dictionary and extracting the observable hydrodynamic data would not only be interesting in itself, but could also provide a window to explicitly check our predictions for the transport coefficients at strong coupling.
Understanding how torsion effects can be incorporated into a `continuous' description could also shed light on how anomalies of this kind are manifest in other low-energy effective theories, such as non-linear sigma models for non-abelian gauge theory. See {\em e.g.} \cite{Yonekura:2020upo,Lee:2020ojw,Davighi:2020vcm,Hason:2020yqf,Hsin:2022heo,Kobayashi:2023ajk} for recent work in this direction, and real-time prescriptions for hydrodynamic effective actions.\footnote{
When viewed as a theory of light degrees of freedom fluctuating around thermal equilibrium, one can also recast hydrodynamics using a `pion-like' coset construction {\em i.e.} as a non-linear sigma model~\cite{Dubovsky:2011sj}, with dissipation obtained by putting the theory in the complex time contour of the Schwinger--Keldysh formalism. See {\em e.g.} \cite{Dubovsky:2011sj,Liu:2018kfw} for a review, and {\em e.g.} \cite{Glorioso:2017lcn,Nair:2011mk,Haehl:2013hoa} for discussions concerning anomalies in real-time hydrodynamic effective actions.
}

Another interesting avenue is to understand new symmetry structures that arise upon gauging the $\U$ or $\mathbb{Z}_2$ subgroup. It is known, in the case of $\U\times \U$ symmetry with a perturbative mixed anomaly, that the gauging can result in either 2-group global symmetry \cite{Cordova:2018cvg} or non-invertible symmetry \cite{Choi:2022jqy,Cordova:2022ieu}. 
The mixed $\U\times \mathbb{Z}_2$ anomaly can suggest a new pathway to such categorical symmetries, and its connection to hydrodynamics that we explored in this paper might provide a link toward potential observable consequences.

Last but not least, we should highlight the possibility of {\em experimentally observing} the signature of the new mixed anomaly. It is known that the low-energy excitations of a class of materials called Weyl semimetals are governed by the Weyl equation \cite{turner2013beyond}. This class of system exhibits perturbative anomalies, in particular the mixed $\U$-gravitational anomaly \cite{Son:2012bg,Lucas:2016omy}, whose effect has been observed experimentally \cite{Gooth:2017mbd}. Given the simplicity of the microscopic realisation set out in \S \ref{sec:formal-stuff}, it does not seem outside the realm of possibility that the mixed $\U\times \mathbb{Z}_2$ anomaly could be realised in related materials and that its salient feature, the $\mathbb{Z}_4$-quantised transport coefficients, could be measured in a similar experimental setup. 


\section*{Acknowledgements}
We would like to thank I\~naki Garc\'ia-Etxebarria, Nabil Iqbal, Akash Jain, Tin Sulejmanpasic and David Tong for various discussions on this topic, as well as Karl Landsteiner and David Tong for their comments on our manuscript.  
NL is supported by the STFC consolidated grant in Particles, Strings and
  Cosmology number ST/T000708/1 and the Royal Society of London.
  The work of NP is supported by the grant for development of new faculty (Ratchadapiseksomphot fund) and Sci-Super IX 66 004 from Chulalongkorn University and is grateful to the hospitality of Durham University and NORDITA during the course of this work. NL and NP also wish to thank the “Paths to Quantum Field Theory 2023” workshop for hospitality during part of this project.
\appendix

\section{Bordism computation for the mixed anomaly} \label{app:bordism}


In this Appendix, we first give a short review of the Adams Spectral
Sequence (ASS)~\cite{Adams:1958} for computing spin bordism groups in~\ref{sec:bordism-via-ASS}, before presenting our computations for
$\Omega^{\Spin}_d(B\U \times B\Z_2)$, $\Omega^{\Pin^-}_d(B\U)$, and
$\Omega^{\Spinc}_d(B\Z_2)$ in~\ref{sec:comp-with-ASS},
\ref{sec:Pin-bord-BU1}, and \ref{sec:bord-spinc-BZ2},
respectively. For a detailed practical guide to using the ASS in
computing bordism groups, see {\em e.g.}  \cite{beaudry2018guide}
and \cite{Campbell:2017khc}. In the rest of this Section, all
cohomology groups are understood to have $\Z_2$ coefficients.

\subsection{Spin bordism groups via the Adams spectral sequence}
\label{sec:bordism-via-ASS}

A cohomological spectral sequence $( E^{s,t}_r , \dd_r)$
is a sequence of bi-graded Abelian groups $E^{s,t}_r$ (in all our
cases, the gradings $s,t$ are in the range $\N_{\geq 0}$),
$r=2,3,4, \ldots$, together with homomorphisms
\begin{equation}
\dd_r: E^{s,t}_r \to E^{s+r,t-1+r}_r \;.
\end{equation}
These homomorphisms are required to be differentials, meaning they should be nilpotent {\em viz.}
$\dd_r\circ \dd_r = 0$, so that $E^{s,t}_r$ at a fixed $r$ form a layer of
differential cochain complexes. The `next page' $E^{s,t}_{r+1}$ is computed by taking the
homology of the complex $E^{s,t}_r$ with respect to $\dd_r$. The set of groups $E^{s,t}_r$ for a given value of $r$ (for all
bi-gradings $(s,t)$), sometimes denoted simply as $E_r$, are
collectively likened to a page of a book, with $r$ serving as a page
number. The process of computing successive $E_r$ is then likened to turning the pages of
an (infinitely long) book, called the spectral sequence.

Thankfully, at least in all cases where the spectral sequence yields a calculable result, only a finite number of pages is needed.
At a fixed bi-grading $(s,t)$, the page-turning process comes to an end after a
certain number of steps because the entry stabilises:
$E^{s,t}_r = E^{s,t}_R$, $\forall r \geq R$ for some $R$ (which depends on
$s$ and $t$). We conventionally call the stabilised entry as belonging
to the `last page' of the spectral sequence, and denote it by
$E^{s,t}_{\infty}$.

The spectral sequence relevant to our spin bordism computation is a
specific version of the Adams spectral sequence (ASS), used for the
computation of stable homotopy groups in general. It is defined by how
the $E_2$-page is built, and what it `converges' to. Our ASS that
computes $\Omega^{\Spin}_n(BG)$ for a symmetry group $G$ is given by
\begin{equation}
E_2^{s,t} = \ext^{s,t}_{\mathcal{A}(1)} \left( H^{\bullet}(BG), \Z_2 \right) \Longrightarrow \left( \Omega^{\Spin}_{t-s}(BG) \right)^{\wedge}_2\,.
\label{eq:ASS-for-spin-bordism}
\end{equation}
Here $\mathcal{G}^{\wedge}_2$ denotes completion at 2 of an abelian group $\mathcal{G}$ (see {\em e.g.}~\cite[Definition 4.7.12]{beaudry2018guide} for a definition), where
$(\Z)^{\wedge}_2$ is the 2-adic integers and
$\left( \Z_m \right)^{\wedge}_2$ is equal to $\Z_m \otimes \Z_2$, and
$H^{\bullet}(BG)$ is the mod-2 cohomology of $BG$, thought of as a module over
the Steenrod subalgebra $\mathcal{A}(1)$ generated by the lowest degree Steenrod squares $\sq^0$,
$\sq^1$ and $\sq^2$. These Steenrod squares $\sq^i$, defined in general for all $i\geq 0$, are stable
cohomology operations {\em i.e.} natural transformations between
cohomology functors of the form
\begin{equation}
\sq^i : H^{\bullet}(X) \to H^{\bullet+i}(X)\, ,
\end{equation}
such that the following axioms are satisfied:
\begin{enumerate}
\item $\sq^0$ is the identity homomorphism;
\item $\sq^i(x) = 0$ if $i>\abs{x}$, the degree of $x$;
\item $\sq^i(x)= x\cup x$ if $\abs{x}= i$;
\item {\it Cartan formula} $\sq^i( x\cup y) = \sum_{j+k=i} \sq^j(x) \cup \sq^k(y)$ .
\end{enumerate}

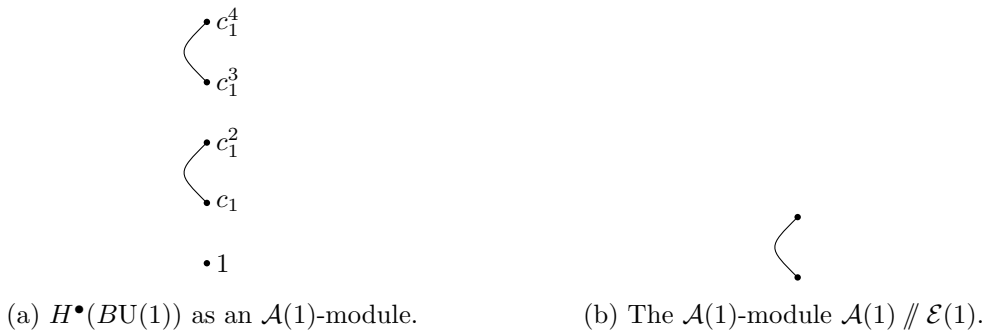
\begin{figure}[h!]
  \centering
  \begin{subfigure}[b]{0.45\textwidth}
    \centering
  \begin{tikzpicture}[scale=0.4]
    \fill (0,0) circle (3pt) node[anchor=west] {$1$};
    \fill (0,2) circle (3pt) node[anchor=west] {$c_1$};
    \fill (0,4) circle (3pt) node[anchor=west] {$c_1^2$};
    \sqtwoL(0,2,black);
    \fill (0,6) circle (3pt) node[anchor=west] {$c_1^3$};
    \fill (0,8) circle (3pt) node[anchor=west] {$c_1^4$};
    \sqtwoL(0,6,black);
  \end{tikzpicture}
  \caption{$H^\bullet(B\U)$ as an $\mathcal{A}(1)$-module.}
  \label{fig:HBU1-module}
\end{subfigure}
\begin{subfigure}[b]{0.45\textwidth}
  \centering
    \begin{tikzpicture}[scale=0.4]
    \fill (0,2) circle (3pt);
    \fill (0,4) circle (3pt);
    \sqtwoL(0,2,black);
      \end{tikzpicture}
  \caption{The $\mathcal{A}(1)$-module $\mathcal{A}(1)\sslash \mathcal{E}(1)$.}
  \label{fig:A1E1}
\end{subfigure}
\caption{Graphical representations of two $\mathcal{A}(1)$-modules.}
\end{figure}

\begin{figure}[h!]
  \centering
  \begin{subfigure}[b]{0.45\textwidth}
    \centering
  \begin{tikzpicture}[scale=0.5]
  \emm(0,1,magenta,\text{\tiny (degree $1$)});
  \end{tikzpicture}
  \caption{$R_0$.}
  \label{fig:R0-module}
\end{subfigure}
\begin{subfigure}[b]{0.45\textwidth}
  \centering
  \begin{tikzpicture}[scale=0.5]
        \fill[teal] (4,3) circle (3pt) node[anchor=north] {\tiny (degree $0$)};
    \fill[teal] (4,4) circle (3pt);
    \fill[teal] (4,5) circle (3pt);
    \fill[teal] (4,6) circle (3pt);
    \fill[teal] (4,7) circle (3pt);
     \fill[teal] (4,8) circle (3pt);
    \fill[teal] (4,9) circle (3pt);
    \sqone(4,3,teal);
    \sqtwoL(4,3,teal);
    \sqone(4,5,teal);
    \sqtwoL(4,6,teal);
    \sqone(4,7,teal);
    \sqtwoR(4,7,teal);
    \draw[dashed,teal] (4,9) -- (4,10);

        \fill[teal] (4+2,3+2) circle (3pt);
    \fill[teal] (4+2,4+2) circle (3pt);
    \fill[teal] (4+2,5+2) circle (3pt);
    \fill[teal] (4+2,6+2) circle (3pt);
    \fill[teal] (4+2,7+2) circle (3pt);
     \fill[teal] (4+2,8+2) circle (3pt);
    \fill[teal] (4+2,9+2) circle (3pt);
    \sqone(4+2,3+2,teal);
    \sqtwoR(4+2,3+2,teal);
    \sqone(4+2,5+2,teal);
    \sqtwoR(4+2,6+2,teal);
    \sqone(4+2,7+2,teal);
    \sqtwoL(4+2,7+2,teal);
    \draw[dashed,teal] (4+2,9+2) -- (4+2,10+2);

    \sqtwoCR(4,4,teal);
      \end{tikzpicture}
  \caption{$R_6$.}
  \label{fig:R6-module}
\end{subfigure}
\caption{Graphical representations of the other two
  $\mathcal{A}(1)$-modules used in the main part of the paper.}
\end{figure}
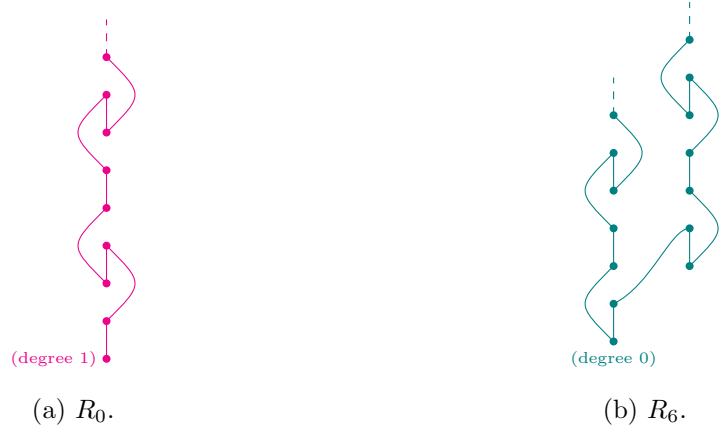

To write down the 2\textsuperscript{nd} page
$E_2^{s,t} = \ext^{s,t}_{\mathcal{A}(1)}(H^{\bullet}(BG),\Z_2)$, we must know 
the $\mathcal{A}(1)$-module structure of $H^{\bullet}(BG)$. This task is aided by the
conventional graphical representation of $\mathcal{A}(1)$-modules, which we now
describe. Firstly, $\Z_2$ summands of $H^{\bullet}(BG)$, considered {\it as a cohomology ring},
are represented by dots on a vertical ladder according to their
degrees in the cohomology ring. To show the actions of the elements of
$\mathcal{A}(1)$ on the module, we draw a vertical line between two dots
separated by 1 degree for the action of $\sq^1$, while a curved line
that jumps by 2 degrees represents $\sq^2$. As an example, consider
$H^{\bullet}(B\U) \cong \Z_2[c_1]$, where $c_1\in H^2(B\U)$ is the mod 2 reduction
of the universal first Chern class. By the axioms and the degree
constraint, we have
\begin{equation}
\sq^1 (c_1) = 0, \qquad \sq^2 (c_1) = c_1 \cup c_1, \qquad \sq^2(c_1 \cup c_1) = 2c_1^3=0, \qquad \dots  \,
\end{equation}
One can then easily represent the lower degrees of $H^{\bullet}(B\U)$ as an
$\mathcal{A}(1)$-module by Fig. \ref{fig:HBU1-module}. We clearly see that
$H^{\bullet}(B\U)$ can be written as a direct sum of more fundamental
$\mathcal{A}(1)$-modules, as
\begin{equation}
H^{\bullet}(B\U) \cong \Z_2 \oplus \Sigma^2 \mathcal{A}(1)\sslash \mathcal{E}(1) \oplus \Sigma^6 \mathcal{A}(1)\sslash \mathcal{E}(1) \oplus \ldots \,,
\end{equation}
where $\mathcal{A}(1)\sslash \mathcal{E}(1)$ is an
$\mathcal{A}(1)$-module represented by Fig. \ref{fig:A1E1} and the suspension
$\Sigma^n$ simply raises the degrees by $n$. Other
$\mathcal{A}(1)$-modules that appear in the computation of
$\Omega^{\Spin}_5(B\U\times B\Z_2)$ in the next Subsection are shown in
Figs. \ref{fig:R0-module} and \ref{fig:R6-module}.

Given the $\mathcal{A}(1)$-module structure of $H^{\bullet}(BG)$, one can then compute
$\ext^{s,t}_{\mathcal{A}(1)}(H^{\bullet}(BG),\Z_2)$ from first principles using tools from
homological algebra as described in detail in {\em e.g.}
Ref. \cite[\S\S 4.4,4.5]{beaudry2018guide}. We will not cover
this here, relying instead on the many results already computed in
\cite{beaudry2018guide} and elsewhere in the literature. Once it
is computed, $\ext^{s,t}_{\mathcal{A}(1)}(H^{\bullet}(BG),\Z_2)$ is represented on a
2d grid called the {\it Adams chart}, with $s$ as the vertical
coordinate and the topological degree $t-s$ as the horizontal
coordinate. Each $\Z_2$ summand is represented by a dot. For example,
$\ext^{s,t}_{\mathcal{A}(1)}(\Z_2,\Z_2)$ is given in the Adams chart form as in
Fig. \ref{fig:extZ2-chart} below. The various lines reflect the fact
that $\ext^{s,t}_{\mathcal{A}(1)}(H^{\bullet}(BG),\Z_2)$ is a module over
$\ext^{s,t}_{\mathcal{A}(1)}(\Z_2,\Z_2)$. Here, a vertical line always expresses
a multiplication by the generator $h_0$ of
$\ext^{1,1}_{\mathcal{A}(1)}(\Z_2,\Z_2)$, while a $45$-degree line represents a
multiplication by the generator $h_1$ of
$\ext^{1,2}_{\mathcal{A}(1)}(\Z_2,\Z_2)$. They also satisfy
\begin{equation}
\dd_r (h_ix) = h_i \dd_r(x), \quad \text{for } i=0,1,
\end{equation}
meaning that $\dd_r$ are $h_0$- and $h_1$-linear.
\begin{figure}[h!]
  \centering
 \includegraphics[scale=0.75]{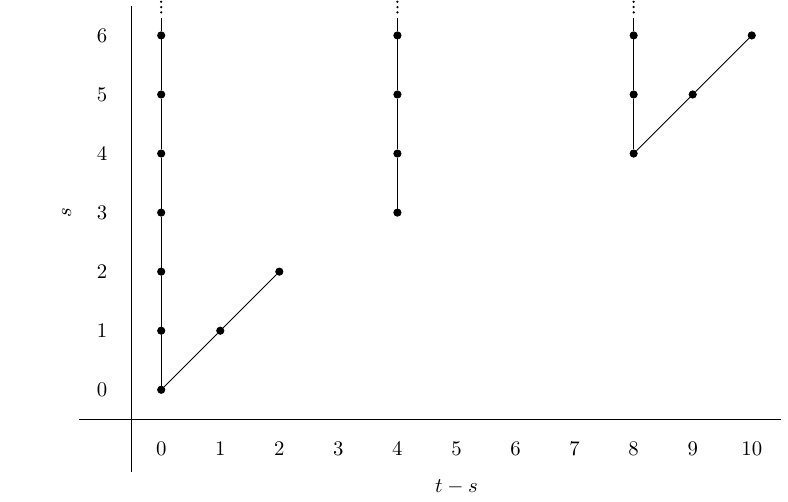} 
  \caption{The Adams chart for $\ext^{s,t}_{\mathcal{A}(1)}(\Z_2,\Z_2)$.}
  \label{fig:extZ2-chart}
\end{figure}
For more examples, we display the Adams charts of
$\ext^{s,t}_{\mathcal{A}(1)}(M,\Z_2)$ for the $\mathcal{A}(1)$-modules
$M= R_0$, $\mathcal{A}(1)\sslash \mathcal{E}(1)$, and $R_6$, that will appear in the computation of
$\Omega^{\Spin}_5(B\U\times B\Z_2)$ in the next Subsection, in
Figs. \ref{fig:Adams-chart-BZ2}, \ref{fig:Adams-chart-A1E1}, and
\ref{fig:Adams-chart-R6}, respectively.

\begin{figure}[h]
  \centering
\begin{subfigure}[b]{0.3\textwidth}
  \centering
  \includegraphics[scale=0.52]{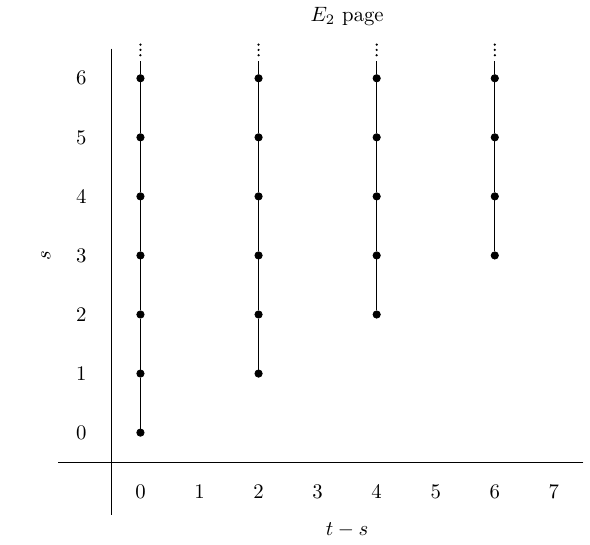}
  \caption{Adams chart for $\ext^{s,t}_{\mathcal{A}(1)}(\mathcal{A}(1)\sslash \mathcal{E}(1),\Z/2)$.}
  \label{fig:Adams-chart-A1E1}
\end{subfigure}
  \begin{subfigure}[b]{0.3\textwidth}
  \centering
  \includegraphics[scale=0.52]{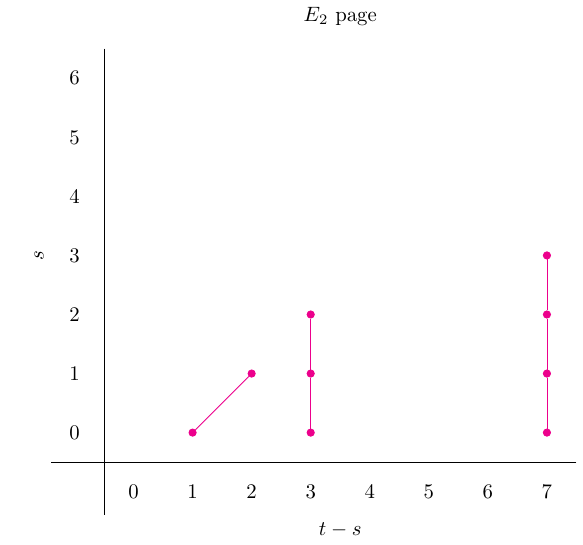}
  \caption{Adams chart for $\ext^{s,t}_{\mathcal{A}(1)}(R_0,\Z/2)$.}
  \label{fig:Adams-chart-BZ2}
\end{subfigure}%
  \begin{subfigure}[b]{0.3\textwidth}
  \centering
  \includegraphics[scale=0.52]{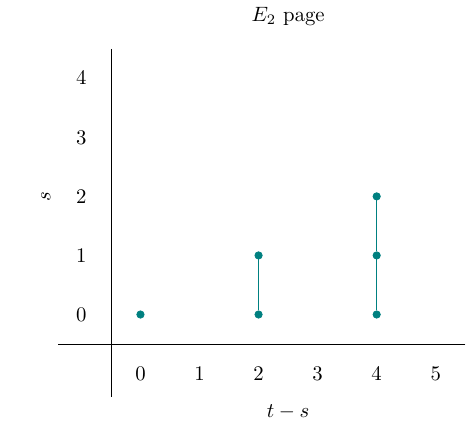}
  \caption{Adams chart for $\ext^{s,t}_{\mathcal{A}(1)}(R_6,\Z/2)$.}
  \label{fig:Adams-chart-R6}
  \end{subfigure}
\caption{Other Adams charts that we need for the computation of $\Omega^{\Spin}_5(B\U\times B\Z_2)$.}
\end{figure}

The ASS $(E^{s,t}_{r}, \dd_r)$ is said to converge to
$\left(\Omega^{\Spin}_n(BG)\right)^{\wedge}_2$, meaning
that there is a filtration $\{ F^{s,n+s}_{\infty}\}_{s\in \N}$ of
$\left(\Omega^{\Spin}_n(BG)\right)^{\wedge}_2$:
\begin{equation}
\left( \Omega^{\Spin}_n(BG) \right)^{\wedge}_2 = F^{0,n}_{\infty} \supseteq F^{1,n+1}_{\infty} \supseteq F^{2,n+2}_{\infty}\supseteq \ldots
\end{equation}
where the quotients of adjacent layers are given by the last page of
the ASS as follows
\begin{equation}
F^{s,n+s}_{\infty}\big/F^{s+1,n+s+1}_{\infty} = E^{s,n+s}_{\infty}\,.
\end{equation}
These relations can be turned into successive short exact sequences
for $F^{0,n}_{\infty}\big / F^{s,n+s}_{\infty}$, $s\geq 1$, which allow us to
obtain $F^{0,n}_{\infty}\big / F^{s+1,n+s+1}_{\infty}$ from
$F^{0,n}_{\infty}\big / F^{s,n+s}_{\infty}$ by solving the extension problems
\begin{equation}
  0 \to E^{s,n+s}_{\infty} \to F^{0,n}_{\infty}\big / F^{s+1,n+s+1}_{\infty} \to F^{0,n}_{\infty}\big / F^{s,n+s}_{\infty} \to 0 \label{eq:Adams-extension-problems}
\end{equation}
with the initial term
$F^{0,n}_{\infty}\big / F^{1,n+1}_{\infty} \cong E^{0,n}_{\infty}$. The final result
$\left( \Omega^{\Spin}_n(BG) \right)^{\wedge}_2$ is given by the inverse limit
of all these quotients:
\begin{equation}
\left( \Omega^{\Spin}_n(BG) \right)^{\wedge}_2\; \cong \;\varprojlim_s F^{0,n}_{\infty}\big / F^{s,n+s}_{\infty}\,.
\end{equation}

Solving the extension problems \eqref{eq:Adams-extension-problems} is
generally hard. However, we are aided by the
$\ext^{\bullet,\bullet}_{\mathcal{A}(1)}(\Z_2,\Z_2)$-module structure on
$E_{\infty}$. It is known that if there is an $h_0$-multiplication
connecting $E^{s,n+s}_{\infty}$ and $E^{s-1,n+s-1}_{\infty}$, the extension
\eqref{eq:Adams-extension-problems} is non-trivial~\cite{beaudry-campbell:2018}. If there are no
exotic extensions,\footnote{Exotic extensions are non-trivial
  extensions that do not come from $h_0$-multiplications.} we can
recover the 2-completion factors in $\Omega^{\Spin}_n(BG)$ from the last
page of the Adams chart, by following the two practical `rules':
\begin{itemize}
\item A finite string of $m\geq 1$ dots connected by $h_0$-lines in the column
  $t-s=n$ corresponds to a factor of $\Z_{2^m}$ in
  $\left(\Omega^{\Spin}_n(BG)\right)^{\wedge}_2$ and ultimately a factor of
  $\Z_{2^m}$ in $\Omega^{\Spin}_n(BG)$.
\item An infinitely tall $h_0$-tower in the column $t-s=n$ corresponds to a factor of
  the $2$-adic integers $\Z^{\wedge}_2$ in
  $\left(\Omega^{\Spin}_n(BG)\right)^{\wedge}_2$, which comes from a
  $\Z$ factor in $\Omega^{\Spin}_n(BG)$.
\end{itemize}

\subsection{Computation of \texorpdfstring{$\Omega^{\Spin}_d(B\U\times B\Z_2)$}{spin bordism groups of BU(1) × Bℤ₂}} \label{sec:comp-with-ASS}

 We are now ready to compute the bordism group
$\Omega^{\Spin}_5(B\U \times B\Z_2)$ using the Adams spectral sequence (ASS)
outlined in the previous Subsection.  The ASS is initialised in this
case on the second page, which is given by the formula:
\begin{equation}
E_2^{s,t} = \ext^{s,t}_{\mathcal{A}(1)} \left( \tilde{H}^{\bullet}(BG),\Z_2 \right) \Rightarrow \left( \tilde{\Omega}^{\Spin}_{t-s}(BG) \right)^{\wedge}_2\;.
\end{equation}
The observant reader will notice the appearance of `tildes' with respect to~\eqref{eq:ASS-for-spin-bordism}, on both cohomology and bordism groups. This is because we here use a variant of the ASS adapted for {\em reduced} bordism groups
$\tilde{\Omega}^{\Spin}_d(BG)$, defined from the natural splitting of
the full bordism group as 
\begin{equation}
  \Omega^{\Spin}_d(BG) \cong \Omega^{\Spin}_d(\text{pt}) \oplus \tilde{\Omega}^{\Spin}_d(BG)\,.
  \label{eq:reduced-bordism-splitting}
\end{equation}
The reduced cohomology groups $\tilde{H}^i(BG)$ are similarly defined by 
\begin{equation}
  H^i(BG) \cong H^i(\text{pt}) \oplus \tilde{H}^i(BG)\,.
  \label{eq:reduced-cohomology-splitting}
\end{equation}
The full bordism groups can be then be assembled from the reduced spin
bordism groups and the well-known result for the spin bordism groups of
a point~\cite{anderson1966spin}, which we here reproduce in Table \ref{tab:spin-bord-pt}.
\begin{table}[h!]
  \centering
  \begin{tabular}{|c|ccccccc|}
    \hline
    $d$ & $~~0~~$ & $~~1~~$ & $~~2~~$ & $~~3~~$ & $~~4~~$ & $~~5~~$ & $~~6~~$\\
    \hline
    $\Omega^{\Spin}_d(\text{pt})$ & $\Z$ & $\Z_2$ & $\Z_2$ & $0$ & $\Z$ & $0$ & $0$\\
    \hline
  \end{tabular}
  \caption{The spin bordism groups of a point~\cite{anderson1966spin}.
  \label{tab:spin-bord-pt}}
\end{table}

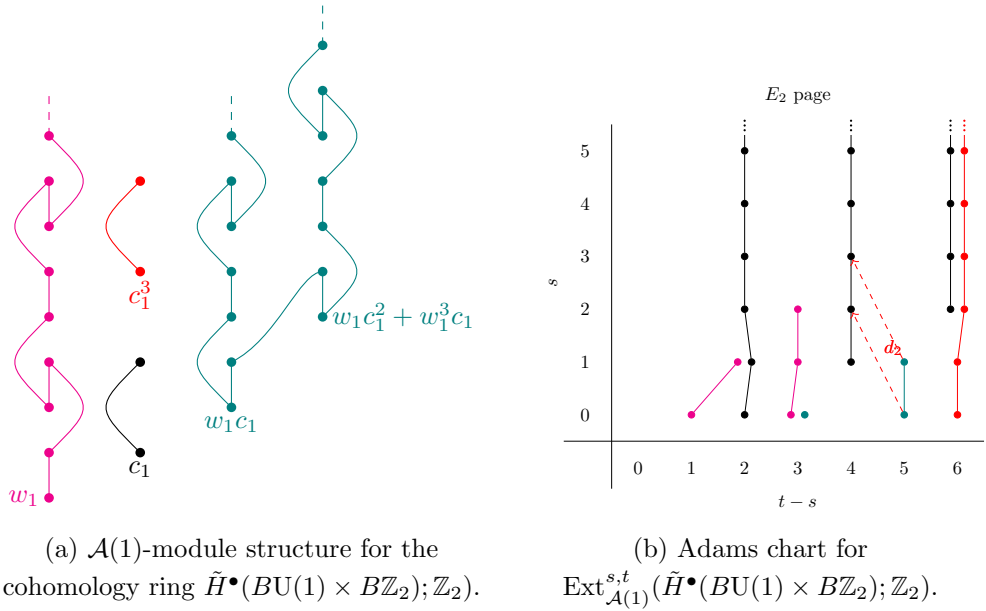
\begin{figure}
\centering
\begin{subfigure}[t]{0.4\textwidth}
    \centering
    \begin{tikzpicture}[scale=0.6]
    \emm(0,1,magenta,w_{1});

    \fill (2,2) circle (3pt) node[anchor=north] {$c_1$};
    \fill (2,4) circle (3pt);
    \sqtwoL(2,2,black);

    \fill[red] (2,6) circle (3pt) node[anchor=north] {$c_1^3$}; \fill[red] (2,8)
    circle (3pt); \sqtwoL(2,6,red);

    \fill[teal] (4,3) circle (3pt) node[anchor=north] {$w_1c_1$};
    \fill[teal] (4,4) circle (3pt);
    \fill[teal] (4,5) circle (3pt);
    \fill[teal] (4,6) circle (3pt);
    \fill[teal] (4,7) circle (3pt);
     \fill[teal] (4,8) circle (3pt);
    \fill[teal] (4,9) circle (3pt);
    \sqone(4,3,teal);
    \sqtwoL(4,3,teal);
    \sqone(4,5,teal);
    \sqtwoL(4,6,teal);
    \sqone(4,7,teal);
    \sqtwoR(4,7,teal);
    \draw[dashed,teal] (4,9) -- (4,10);

        \fill[teal] (4+2,3+2) circle (3pt) node[anchor=west] {$w_1c_1^2+w_1^3c_1$};
    \fill[teal] (4+2,4+2) circle (3pt);
    \fill[teal] (4+2,5+2) circle (3pt);
    \fill[teal] (4+2,6+2) circle (3pt);
    \fill[teal] (4+2,7+2) circle (3pt);
     \fill[teal] (4+2,8+2) circle (3pt);
    \fill[teal] (4+2,9+2) circle (3pt);
    \sqone(4+2,3+2,teal);
    \sqtwoR(4+2,3+2,teal);
    \sqone(4+2,5+2,teal);
    \sqtwoR(4+2,6+2,teal);
    \sqone(4+2,7+2,teal);
    \sqtwoL(4+2,7+2,teal);
    \draw[dashed,teal] (4+2,9+2) -- (4+2,10+2);

    \sqtwoCR(4,4,teal);
  \end{tikzpicture}
  \caption{$\A(1)$-module structure for the cohomology ring $\tilde{H}^{\bullet}(B\U \times B\Z_2);\Z_2)$.\label{fig:A1-mod-structure}}
\end{subfigure}
\begin{subfigure}[t]{0.4\textwidth}
  \centering\includegraphics[scale=0.7]{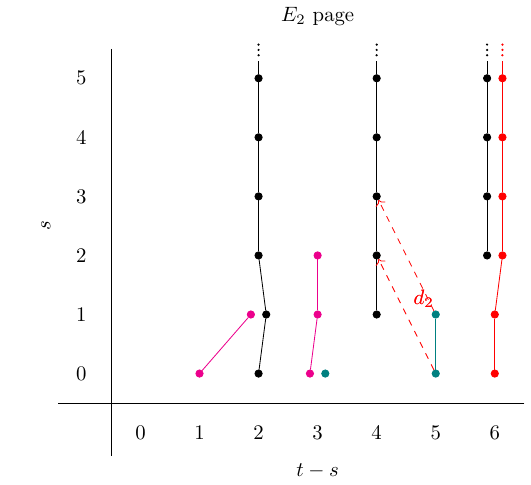}
  \caption{Adams chart for $\ext^{s,t}_{\mathcal{A}(1)}(\tilde{H}^{\bullet}(B\U \times B\Z_2);\Z_2)$.}
  \label{fig:Adams-chart-BZ2-U1}
\end{subfigure}
\caption{Module structure and Adams chart relevant to the new mod 4 anomaly that is central to this paper, and in particular to the fractional hydrodynamic transport constraints that we derive in \S \ref{sec:transport}. \label{fig:steps-for-ASS-of-BZ2-U1}}
\end{figure}

We are interested in the case $G=\U \times \Z_2$. First of all,
we need the structure of the cohomology ring
$H^{\bullet}(B\U \times B\Z_2)$ as an $\A(1)$-module, which is easy to piece
together: it is a classical result that
\begin{equation}
H^{\bullet}(B\Z_2) \cong \Z_2[w_1],\qquad H^{\bullet}(B\U) \cong \Z_2[c_1]\;,
\end{equation}
where $w_1$ and $c_1$ are the universal first Stiefel--Whitney class of $\Z_2 \cong O(2)$
and the universal first Chern class of $\U$ respectively, and so by K\"unneth's theorem we get
\begin{equation}
H^{\bullet}(B\U\times B\Z_2) \cong \Z_2[w_1,c_1]\;.
\end{equation}
Its $\A(1)$-module structure is shown in
Fig. \ref{fig:A1-mod-structure}.

Thus, up to degree 6, the $\mathcal{A}(1)$-module
$\tilde{H}^{\bullet}(B\Z_2\times B\U)$ can be written as the direct sum
\begin{equation}
  \tilde{H}^{\bullet}(B\U \times B\Z_2) \cong {\color{magenta}R_0} \oplus \Sigma^2\mathcal{A}(1)\sslash \mathcal{E}(1) \oplus {\color{red} \Sigma^6\mathcal{A}(1)\sslash \mathcal{E}(1)} \oplus {\color{teal}\Sigma^3 R_6} \oplus \ldots
  \label{eq:A1mod-HBZ2BU}
\end{equation}
adopting the notation of Ref~\cite{beaudry2018guide}, and where
we coordinate the colours between Fig. \ref{fig:A1-mod-structure} and
Eq. \eqref{eq:A1mod-HBZ2BU}. The module $R_0$ is nothing but
$\tilde{H}^{\bullet}(B\Z_2)$ whose Adams chart is known \cite[\S
A.5]{Davighi:2022icj}, already shown in
Fig. \ref{fig:Adams-chart-BZ2}.  Similarly, we also know the Adams
chart for the $\mathcal{A}(1)$-modules
$\mathcal{A}(1)\sslash \mathcal{E}(1)$ and $R_6$, as shown in
Figs. \ref{fig:Adams-chart-A1E1} and \ref{fig:Adams-chart-R6},
respectively ({\it c.f.} \cite[Figs. 22,
41]{beaudry2018guide}). 

We then combine these results together to obtain the Adams chart for
$\tilde{H}^{\bullet}(B\U \times B\Z_2)$, as shown in
Fig. \ref{fig:Adams-chart-BZ2-U1}.
To turn the page, we know that the differentials $\dd_2$ from the column $t-s=5$ must be trivial. This can be proven by contradiction, as follows. Let us denote the generator of the $h_0$-tower in
topological degree $t-s=4$ by $x$, and the generator of the truncated
$h_0$-tower in topological degree $t-s=5$ by $y$, with $h_0^2y=0$. If
$\dd_2$ were non-trivial, then we would have $\dd_2 y = h_0x$. Multiplying
by $h_0^2$ and using the fact that $\dd_2$ and $h_0$ commute, we get
$h_0^3x  = h_0^2\dd_2y = \dd_2 (h_0^2y) = \dd_2(0)=0$,
which is a contradiction. The same argument also shows that the
differentials $\dd_r$ on the $r$\textsuperscript{th} page of the
spectral sequence from topological degrees $t-s=3,5$ are all
trivial. The differentials from topological degrees $t-s=1,2,4,6$ are
trivial simply for degree reasons.

Knowing these differentials, and knowing that there is no
odd torsion, we can read off the reduced spin bordism groups of
$B\U \times B\Z_2$ from the Adams chart \ref{fig:Adams-chart-BZ2-U1} by applying the two practical `rules' given in~\ref{sec:bordism-via-ASS},
whence the full spin bordism groups can be recovered
through \eqref{eq:reduced-bordism-splitting} and Table
\ref{tab:spin-bord-pt}. We summarise all the bordism groups in degrees
0 through 6, as recorded in Table \ref{fig:spin-bord-BU-BZ2-text} in
the main text. In particular, in degree 5 we obtain the $\Z_4$
reported in Eq.~(\ref{eq:Z4}) that is the main object of physical interest in this paper.

\subsection{Computation of \texorpdfstring{$\Omega^{\Pin^-}_d(B\U)$}{pin- bordism groups of BU(1)}}
\label{sec:Pin-bord-BU1}

To compute the bordism groups for the tangential structure
$\Pin^- \times B\U$, we have to use a slightly different version of the ASS
to the one discussed above. A similar strategy was used to compute spin-$\UU(2)$ bordism groups in~\cite{Davighi:2020bvi}. Applying the Anderson--Brown--Peterson
Theorem \cite{ABP:1967} to the weak equivalence between two
Madsen--Tillmann spectra,
$MT\Pin^- \simeq MT\Spin \wedge \Sigma^{-1} M\O(1)$, where $MTH$ denotes the
Madsen--Tillmann spectrum of $H$ and $M\O(1)$ is the Thom space of
$B\O(1)$, we can show that there is an ASS converging to
$\Omega^{\Pin^-}_d(B\U)$ given by
\begin{equation}
E^{s,t}_2 = \ext^{s,t}_{\mathcal{A}(1)} \left( \Sigma^{-1} \tilde{H}^{\bullet}(M\O(1))\otimes H^{\bullet}(B\U), \Z_2 \right) \Longrightarrow \left(\Omega^{\Pin^-}_{t-s} (B\U)\right)^{\wedge}_2\;.
\end{equation}

The mod 2 cohomology ring $\tilde{H}^{\bullet}(M\O(1))\otimes H^{\bullet} (B\U)$ is given by
\begin{equation}
\tilde{H}^{\bullet}(M\O(1))\otimes H^{\bullet}(B\U) \cong \Z_2 [w_1,c_1] \left\{ U \right\}
\end{equation}
where $U \in \tilde{H}^1(M\O(1))$ is the Thom class. This means a
generic generator of this ring takes the form $c_1^n w_1^m U$.  The
action of the Steenrod subalgebra $\mathcal{A}(1)$ on this ring as an
$\mathcal{A}(1)$-module is standard: the action on $w_1$ and $c_1$ was already
spelled out in Appendix \ref{sec:comp-with-ASS}, while the additional
action on $U$ is given by
\begin{equation}
\sq^1 U = w_1 U, \qquad \sq^2 U = 0\,.
\end{equation}
From this information, the $\mathcal{A}(1)$-module structure of
$\Sigma^{-1} \tilde{H}^{\bullet}(M\O(1))\otimes H^{\bullet}(B\U)$ can be easily inferred:
\begin{equation} \label{eq:module-structure-Pinminus-example}
\Sigma^{-1} \tilde{H}^{\bullet}(M\O(1))\otimes H^{\bullet}(B\U) \cong {\color{magenta}\Sigma^{-1}R_0} \oplus {\color{teal}\Sigma^2R_6} \oplus \ldots
\end{equation}
as shown in Fig. \ref{fig:A1-mod-structure-Pin-}, where the modules
$R_0$ and $R_6$ are the same as in the previous Appendices. In the diagram $U_0$ denotes the virtual Thom class obtained by shifting $U$ down by one degree -- a consequence of the suspension shift $\Sigma^{-1}$ appearing in Eq.~\eqref{eq:module-structure-Pinminus-example} {\em etc}. The
corresponding Adams chart for $E^{s,t}_2$ is shown in
Fig. \ref{fig:Adams-chart-Pin--U1}, whence we can read off the
$\Pin^-$ bordism groups of $B\U$ from degrees $0$ through $5$,
recorded in Table \ref{tab:Pin--bord-U1}.
\begin{table}[h!]
  \centering
  \begin{tabular}{|c|cccccc|}
    \hline
    $d$ & $~~0~~$ & $~~1~~$ & $~~2~~$ & $~~3~~$ & $~~4~~$ & $~~5~~$\\
    \hline
    $\Omega^{\Pin^-}_d(B\U)$ & $\Z_2$ & $\Z_2$ & $\Z_2 \times \Z_8$ & $0$ & $\Z_4$ & $0$\\
    \hline
  \end{tabular}
  \caption{The $\Pin^-$ bordism groups of $B\U$. 
  \label{tab:Pin--bord-U1}}
\end{table}
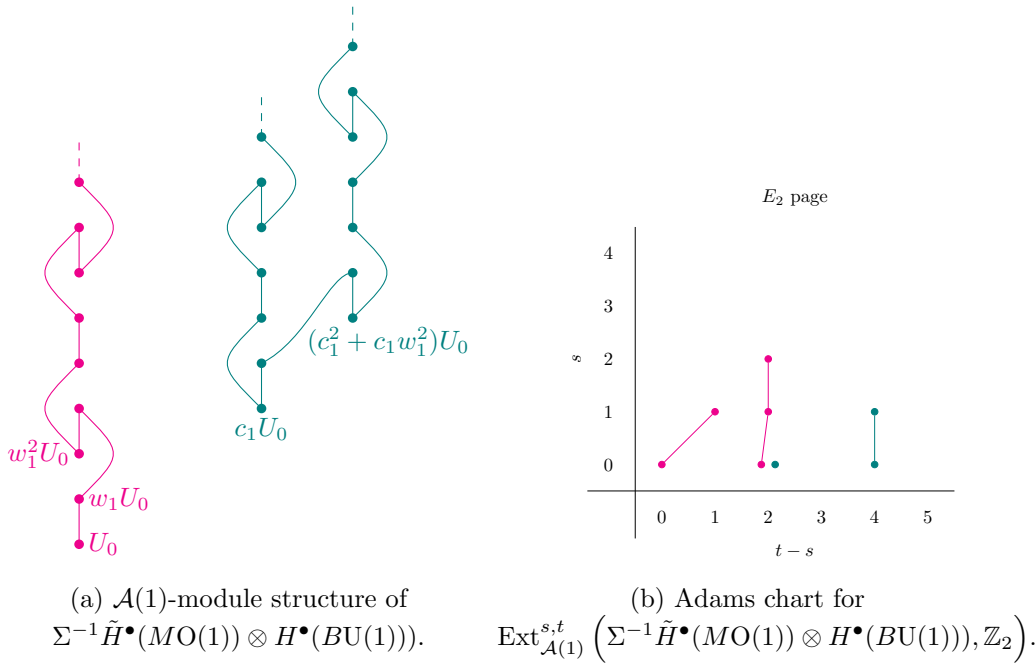
\begin{figure}[h!]
\centering
\begin{subfigure}[t]{0.4\textwidth}
    \centering
    \begin{tikzpicture}[scale=0.6]
  \fill[magenta] (0,0) circle (3pt) node[anchor=west] {$U_0$};
  \fill[magenta] (0,1) circle (3pt) node[anchor=west] {$w_1U_0$};
  \fill[magenta] (0,2) circle (3pt) node[anchor=east] {$w_1^2U_0$};
  \fill[magenta] (0,3) circle (3pt);
  \fill[magenta] (0,4) circle (3pt);
  \fill[magenta] (0,5) circle (3pt);
  \fill[magenta] (0,6) circle (3pt);
  \fill[magenta] (0,7) circle (3pt);
  \fill[magenta] (0,8) circle (3pt);

  \sqone (0,0,magenta);
  \sqtwoR (0,1,magenta);
  \sqone (0,2,magenta);
  \sqtwoL (0,2,magenta);
  \sqone (0,4,magenta);
  \sqtwoL (0,5,magenta);
  \sqone (0,6,magenta);
  \sqtwoR (0,6,magenta);

  \draw[dashed,magenta] (0,8) -- (0,9);

    \fill[teal] (4,3) circle (3pt) node[anchor=north] {$c_1 U_0$};
    \fill[teal] (4,4) circle (3pt);
    \fill[teal] (4,5) circle (3pt);
    \fill[teal] (4,6) circle (3pt);
    \fill[teal] (4,7) circle (3pt);
     \fill[teal] (4,8) circle (3pt);
    \fill[teal] (4,9) circle (3pt);
    \sqone(4,3,teal);
    \sqtwoL(4,3,teal);
    \sqone(4,5,teal);
    \sqtwoL(4,6,teal);
    \sqone(4,7,teal);
    \sqtwoR(4,7,teal);
    \draw[dashed,teal] (4,9) -- (4,10);

        \fill[teal] (4+2,3+2) circle (3pt) node[anchor=north] {$~~~~~~~~(c_1^2 + c_1w_1^2)U_0$};
    \fill[teal] (4+2,4+2) circle (3pt);
    \fill[teal] (4+2,5+2) circle (3pt);
    \fill[teal] (4+2,6+2) circle (3pt);
    \fill[teal] (4+2,7+2) circle (3pt);
     \fill[teal] (4+2,8+2) circle (3pt);
    \fill[teal] (4+2,9+2) circle (3pt);
    \sqone(4+2,3+2,teal);
    \sqtwoR(4+2,3+2,teal);
    \sqone(4+2,5+2,teal);
    \sqtwoR(4+2,6+2,teal);
    \sqone(4+2,7+2,teal);
    \sqtwoL(4+2,7+2,teal);
    \draw[dashed,teal] (4+2,9+2) -- (4+2,10+2);

    \sqtwoCR(4,4,teal);
  \end{tikzpicture}
  \caption{$\A(1)$-module structure of $\Sigma^{-1}\tilde{H}^{\bullet}(M\O(1))\otimes H^{\bullet}(B\U))$.\label{fig:A1-mod-structure-Pin-}}
\end{subfigure}
\begin{subfigure}[t]{0.4\textwidth}
  \centering\includegraphics[scale=0.7]{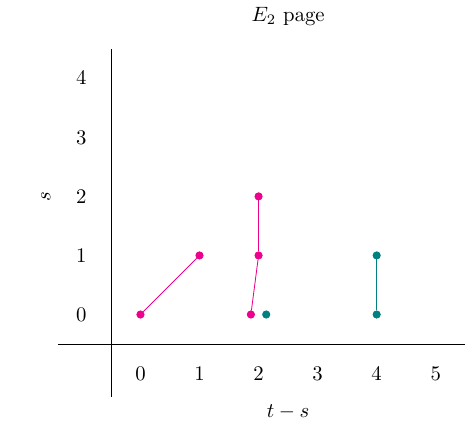}
  \caption{Adams chart for $\ext^{s,t}_{\mathcal{A}(1)}\left(\Sigma^{-1}\tilde{H}^{\bullet}(M\O(1))\otimes H^{\bullet}(B\U)),\Z_2\right)$.}
  \label{fig:Adams-chart-Pin--U1}
\end{subfigure}
\caption{Module structure and Adams chart for $\Pin^- \times \U$ symmetry type, which we believe is relevant for anomaly matching in a domain wall theory obtained by breaking the $\Z_2$ symmetry in the original 4d $\Spin \times \U \times \Z_2$ theory -- see \S \ref{sec:conclusion} of the main text.
\label{fig:steps-for-ASS-of-Pin-bordism}}
\end{figure}

\subsection{Computation of \texorpdfstring{$\Omega^{\Spinc}_d(B\Z_2)$}{spinc bordism groups of  Bℤ₂}}
\label{sec:bord-spinc-BZ2}

Finally, the method we here employ to compute the $\Spinc$ bordism groups of $B\Z_2$
parallels our computation of the $\Pin^-$ bordism groups of $B\U$ in
the previous Subsection. (We stress that these bordism groups have been calculated previously by other means, in work of Bahri and Gilkey~\cite{bahri1987eta}.) The weak equivalence of spectra
\begin{equation}
MT\Spinc \cong MT\Spin \wedge \Sigma^{-2}M\U,
\end{equation}
gives rise to the following ASS:
\begin{equation}
  E^{s,t}_2 = \ext^{s,t}_{\mathcal{A}(1)} \left( \Sigma^{-2}\tilde{H}^{\bullet}(M\U)\otimes H^{\bullet}(B\Z_2),\Z_2 \right) \Longrightarrow \left(\Omega^{\Spinc}_{t-s}(B\Z_2)\right)^{\wedge}_2.
\end{equation}
The cohomology ring
$\Sigma^{-2}\tilde{H}^{\bullet}(M\U)\otimes H^{\bullet}(B\Z_2)$ is known to be
\begin{equation}
\Sigma^{-2}\tilde{H}^{\bullet}(M\U)\otimes H^{\bullet}(B\Z/2) \cong \Z_2[c_1,w_1]\left\{ U_0 \right\},
\end{equation}
where, as before, $c_1$ and $w_1$ denote the universal first Chern
class and the universal first Stiefel--Whitney class associated with
$\U$ and $\Z_2$, respectively. Note that the (virtual) Thom class
$U_0$ is again in degree $0$, indicated by the subscript, which is this time shifted down {\em two} degrees because of the
suspension $\Sigma^{-2}$. 

\begin{figure}[h!]
\centering
\begin{subfigure}[t]{0.3\textwidth}
  \centering
    \begin{tikzpicture}[scale=0.5]
      \fill (0,0) circle (3pt) node[anchor=north] {$U_0$};
    \fill (0,2) circle (3pt) node[anchor=east] {$c_1 U_0$};
    \sqtwoL(0,0,black);

    \fill[red] (0,4) circle (3pt) node[anchor=north] {$c_1^3 U_0$};
    \fill[red] (0,6) circle (3pt); \sqtwoL(0,4,red);

    \fill[brown] (0,8) circle (3pt);
    \fill[brown] (0,10) circle (3pt);
    \sqtwoL(0,8,brown);

    \fill[teal] (4-2,3-2) circle (3pt) node[anchor=north] {${\scriptstyle w_1 U_0}$};
    \fill[teal] (4-2,4-2) circle (3pt);
    \fill[teal] (4-2,5-2) circle (3pt);
    \fill[teal] (4-2,6-2) circle (3pt);
    \fill[teal] (4-2,7-2) circle (3pt);
     \fill[teal] (4-2,8-2) circle (3pt);
    \fill[teal] (4-2,9-2) circle (3pt);
    \sqone(4-2,3-2,teal);
    \sqtwoL(4-2,3-2,teal);
    \sqone(4-2,5-2,teal);
    \sqtwoL(4-2,6-2,teal);
    \sqone(4-2,7-2,teal);
    \sqtwoR(4-2,7-2,teal);
    \draw[dashed,teal] (4-2,9-2) -- (4-2,10-2);

        \fill[teal] (4,3) circle (3pt) node[anchor=north] {$~~~~~{\scriptstyle (w_1^3+c_1w_1)U_0}$};
    \fill[teal] (4,4) circle (3pt);
    \fill[teal] (4,5) circle (3pt);
    \fill[teal] (4,6) circle (3pt);
    \fill[teal] (4,7) circle (3pt);
     \fill[teal] (4,8) circle (3pt);
    \fill[teal] (4,9) circle (3pt);
    \sqone(4,3,teal);
    \sqtwoR(4,3,teal);
    \sqone(4,5,teal);
    \sqtwoR(4,6,teal);
    \sqone(4,7,teal);
    \sqtwoL(4,7,teal);
    \draw[dashed,teal] (4,9) -- (4,10);

    \sqtwoCR(4-2,4-2,teal);
   
     \fill[blue] (4+2,3+2) circle (3pt) node[anchor=north] {${\scriptstyle w_1 c_1^2 U_0}$};
    \fill[blue] (4+2,4+2) circle (3pt);
    \fill[blue] (4+2,5+2) circle (3pt);
    \fill[blue] (4+2,6+2) circle (3pt);
    \fill[blue] (4+2,7+2) circle (3pt);
     \fill[blue] (4+2,8+2) circle (3pt);
    \fill[blue] (4+2,9+2) circle (3pt);
    \sqone(4+2,3+2,blue);
    \sqtwoL(4+2,3+2,blue);
    \sqone(4+2,5+2,blue);
    \sqtwoL(4+2,6+2,blue);
    \sqone(4+2,7+2,blue);
    \sqtwoR(4+2,7+2,blue);
    \draw[dashed,blue] (4+2,9+2) -- (4+2,10+2);

        \fill[blue] (4+4,3+4) circle (3pt) node[anchor=north] {$~~~~{\scriptstyle (w_1^3+c_1w_1)c_1^2U_0}$};
    \fill[blue] (4+4,4+4) circle (3pt);
    \fill[blue] (4+4,5+4) circle (3pt);
    \fill[blue] (4+4,6+4) circle (3pt);
    \fill[blue] (4+4,7+4) circle (3pt);
     \fill[blue] (4+4,8+4) circle (3pt);
    \fill[blue] (4+4,9+4) circle (3pt);
    \sqone(4+4,3+4,blue);
    \sqtwoR(4+4,3+4,blue);
    \sqone(4+4,5+4,blue);
    \sqtwoR(4+4,6+4,blue);
    \sqone(4+4,7+4,blue);
    \sqtwoL(4+4,7+4,blue);
    \draw[dashed,blue] (4+4,9+4) -- (4+4,10+4);

    \sqtwoCR(4+2,4+2,blue);
  \end{tikzpicture}
  \caption{The $\mathcal{A}(1)$-module structure of $\Sigma^{-2}\tilde{H}^{\bullet}(M\U)\otimes H^{\bullet}(B\Z_2)$ \label{fig:A1-spinc-Z2}}
\end{subfigure}
\hspace{0.1\textwidth}
\begin{subfigure}[t]{0.3\textwidth}
  \centering
  \includegraphics[scale=0.7]{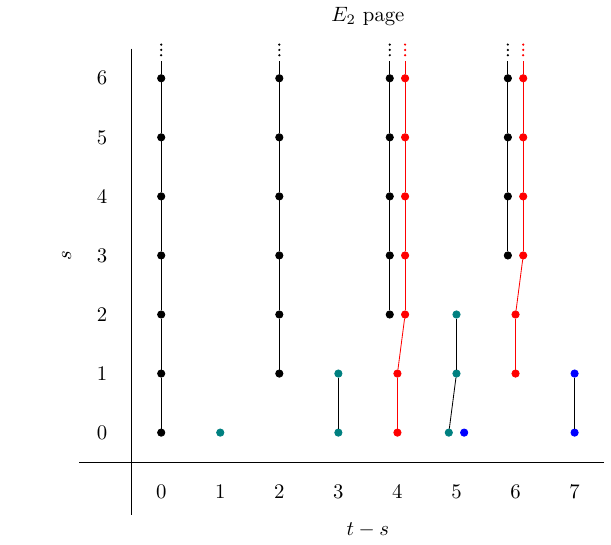}
  \caption{The Adams chart for $\ext^{s,t}_{\mathcal{A}(1)} \left( \Sigma^{-2}\tilde{H}^{\bullet}(M\U)\otimes H^{\bullet}(B\Z_2),\Z_2 \right)$ \label{fig:Adams-spinc-Z2}}
\end{subfigure}
\hspace{0.1\textwidth}
\caption{Module structure and Adams chart relevant for the $\Spinc$ variant of our symmetry type.
\label{fig:steps-for-ASS-of-spinc-bordism}}
\end{figure}
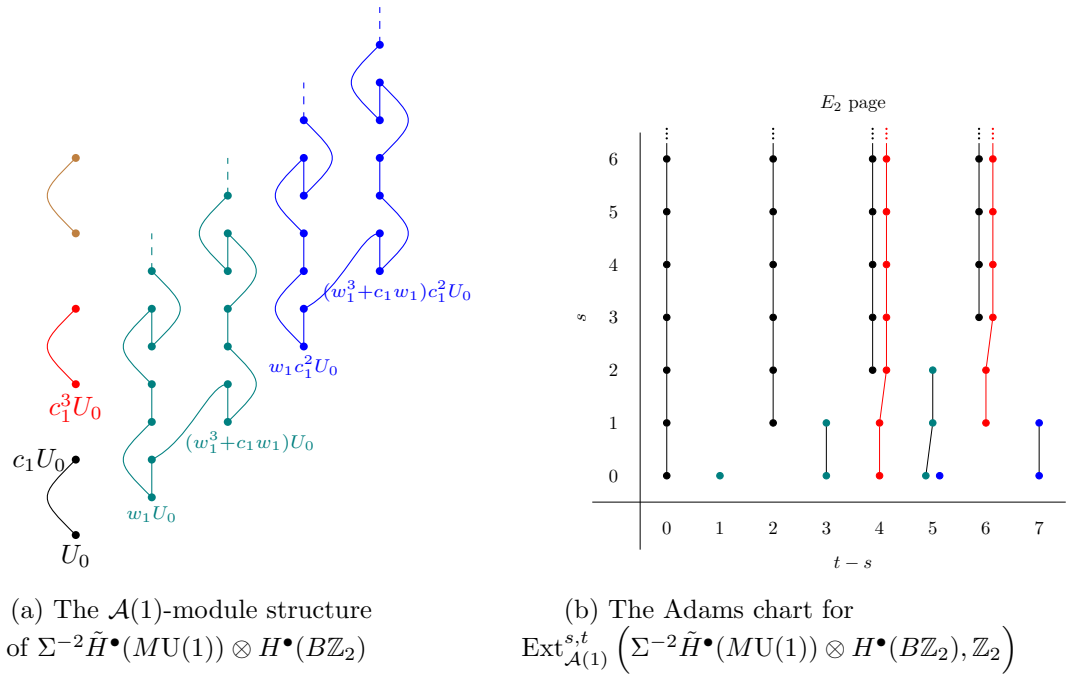

The action of the Steenrod square on $U_0$ is given
by
\begin{equation}
\sq^1 U_0 = 0, \qquad \sq^2 U_0 = c_1U_0. 
\end{equation}
This is sufficient to determine the $\mathcal{A}(1)$-module structure, which we show
in Fig. \ref{fig:A1-spinc-Z2}, with the associated Adams chart of the
Ext-functor from this module to $\Z_2$ given in
Fig. \ref{fig:Adams-spinc-Z2}. We can read off the bordism groups $\Omega^{\Spinc}_d(B\Z_2)$ in degrees
$d=0$ through $6$ directly from the Adams chart, as recorded in Table
\ref{tab:Spinc--bord-Z2}, matching the results of~\cite{bahri1987eta}. In particular, the 5\textsuperscript{th}
$\Spinc$ bordism group of $B\Z_2$ is given by
\begin{equation}
\Omega^{\Spinc}_5(B\Z_2) \cong \Z_8 \times \Z_2.
\end{equation}

\begin{table}[h!]
  \centering
  \begin{tabular}{|c|ccccccc|}
    \hline
    $d$ & $~~0~~$ & $~~1~~$ & $~~2~~$ & $~~3~~$ & $~~4~~$ & $~~5~~$ & $~~6~~$\\
    \hline
    $\Omega^{\Spinc}_d(B\Z_2)$ & $\Z$ & $\Z_2$ & $\Z$ & $\Z_4$ & $\Z^2$ & $\Z_8 \times \Z_2$ & $\Z^2$\\
    \hline
  \end{tabular}
  \caption{The $\Spinc$ bordism groups of $B\Z_2$. 
  \label{tab:Spinc--bord-Z2}}
\end{table}

\bibliographystyle{JHEP.bst}

\bibliography{references}

\end{document}